\documentclass[useAMS,usenatbib]{mn2e}
\usepackage[T1]{fontenc}
\usepackage{aecompl}
\usepackage{graphicx}
\usepackage{times}
\usepackage{booktabs}
\usepackage{subfig}
\usepackage{journals}
\usepackage{color}

\definecolor{bleuf}{rgb}{0.12,0.1,0.6}
\definecolor{grey}{rgb}{0.4,0.6,0.6}
\definecolor{vert}{RGB}{49,148,31}
\definecolor{bleu}{RGB}{143,189,204}

\def\bpar{$b_\parallel$}
\def\bperp{$b_\bot$}

\title[How well does the FoF algorithm recover group properties?]
{How well does the Friends-of-Friends algorithm recover
group properties from galaxy catalogs limited in both distance and luminosity?}
\author[M. Duarte \& G.A. Mamon]{Manuel Duarte$^1$\thanks{E-mail: duarte@iap.fr}, Gary A. Mamon$^1$\thanks{gam@iap.fr}
\\
$^{1}$ Institut d'Astrophysique de Paris, Paris, France (UMR 7095: CNRS \&
UPMC)}

\begin{document}

\date{Accepted?? Received??; in original form??}


\pubyear{2014}

\maketitle

\label{firstpage}

\begin{abstract}
The specific star formation rates of galaxies are
influenced both by their mass and by their environment.
Moreover, the mass function of groups and clusters serves as a powerful
cosmological tool. It is thus important to quantify the accuracy to which
group properties are extracted from redshift surveys.
We test here the Friends-of-Friends (FoF) grouping
algorithm, which depends on two  linking lengths (LLs), plane-of-sky and
line-of-sight (LOS), normalized to the mean nearest neighbor separation of field galaxies.
We argue, on theoretical grounds 
that LLs should be
$b_\perp\simeq 0.11$, and $b_\parallel \approx 1.3$
 to recover 95\% of all galaxies with projected radii
within the virial radius $r_{200}$ and
95\% of the galaxies along the LOS\@.
We then predict that 80 to
90\% of the galaxies in FoF
groups should lie within their
parent real-space groups (RSGs), defined within their virial
spheres.
We test the FoF extraction for 16$\times$16 pairs of  LLs,
using subsamples of galaxies, doubly
complete in distance and luminosity, of a flux-limited mock SDSS galaxy catalog.
%
We find that massive RSGs are more prone to fragmentation, while
the fragments typically have low estimated mass, with typically 30\% of
groups of low
and intermediate estimated mass being fragments. Group merging
rises drastically with estimated mass.
For groups of 3 or more galaxies,
galaxy completeness and reliability are both typically better than 80\%
(after discarding the fragments).
Estimated masses of extracted groups are biased low, by up to a factor 4 at low
richness, while the inefficiency of mass estimation improves from 0.85 dex
to 0.2 dex when moving from low to high multiplicity groups.
The optimal LLs depend on the scientific goal for the group catalog.
We propose $b_\perp\simeq 0.07$, with $b_\parallel\simeq 1.1$ for
studies of environmental effects,
$b_\parallel \simeq 2.5$ for
cosmographic studies and
$b_\parallel \simeq 5$ for followups of individual groups.
%
\end{abstract}


\section{Introduction}
\label{sec:intro}
Galaxies are very rarely isolated:
most live in pairs, groups and clusters of increasing richness and
mass, with mean nearest neighbor separations only one or two orders of
magnitude
greater than their sizes (in contrast to stars within galaxies).
The properties of galaxies are thus expected to be affected by their
\emph{global environment}, the mass of the group in which they reside, and
by their \emph{local environment}, the position they sit within their group.
For example, their specific star formation rate (SSFR) is expected to be
quenched by tidal stripping of their outer gaseous reservoirs by their
group's gravitational potential \citep*{LTC80} and by ram pressure stripping
of these reservoirs and possibly their interstellar gas by the intra-group
gas \citep{GG72}. On the other hand, galaxy collisions and mergers should
trigger bursts of star formation \citep{JW85}, which should later deplete
the galaxies of their gas for subsequent star formation. The respective
roles of these physical processes are still unclear, hence it is important
to probe the global and local environments of galaxies to which models of
galaxy formation can be confronted.

Analyses of the effects of the group environment on the SSFR of galaxies
have led to somewhat discrepant analyses.~\cite{Peng+10} found that only
galaxies of low stellar mass have their SSFR modulated by the environment,
while~\cite{vonderLinden+10} find that the SSFR of high stellar mass
galaxies are also somewhat modulated by their environment. The difference
between these two studies is the lack of distinction between local and
global environments by \citeauthor{Peng+10}. But since it is notoriously
difficult to properly define environment from redshift space catalogs
\citep*{MFW93}, one should strive towards optimal measures of galaxy groups.

Massive groups (i.e., clusters) are also useful as cosmological and physics
tools. For example, the evolution of the cluster mass function is a powerful
diagnostic of cosmological parameters, including dark energy \citep{WS98}.

The extraction of group catalogs from redshift-space data is difficult for
several reasons:
\begin{enumerate}
\item It is intrinsically difficult to characterize systems of a few objects
    (galaxies).
\item The local environment requires an accurate definition of the group
    center,\footnote{The group center is also essential in all studies where
    groups are stacked.} which is also difficult for low-multiplicity
    systems.

\item The Hubble flow creates redshift distortions \citep{Jackson72} that cause galaxies
    within their virial spheres in real space to extend in redshift space by
    $\kappa \eta \sqrt{\Delta/2} \simeq 10-20$ virial radii along the line
    of sight (LOS), where $\kappa \simeq 2-3$ is the number of group
    velocity dispersions that one is studying, $\eta = \sigma_v/v_v \simeq
    0.65$ \citep*{MM07,MBB13} is the group velocity dispersion in units of
    the circular velocity at the virial radius, and $\Delta \simeq 100-200$
    is the mean overdensity at the virial radius relative to the critical
    density of the Universe (see \citealp{Eke+04}; \citealp*{MBM10}). Such elongated groups
    along the LOS risk being confused with other foreground or background
    groups along the same LOS, situated with $\pm 10-20$ virial radii, i.e.\
    typically $10-20$ Mpc. In other words, different groups in real space
    risk being merged, while galaxies found in the group in redshift space
    may not lie within the virial sphere of the group in real space, leading
    to unreliable galaxy membership.
\end{enumerate}

The most popular, and perhaps simplest algorithm is the
\emph{Friends-of-Friends} (FoF) percolation method, which as implied by its
name, puts into a single group all galaxies linked in pairs according to
their separations along the LOS or on the plane-of-sky (POS).

Grouping algorithms are not limited to the FoF technique.~\cite{MDNC02} have
added Delaunay triangulation to Voronoi percolation. Moreover, several
Bayesian methods have been recently developed, taking into account our
\emph{a priori}s, such as assuming NFW models \citep*{NFW96} for the number
and mass density profiles of groups, to conform with the density profiles
measured in $\Lambda$CDM halos \citep{NFW96}. For example,~\cite{YMvdBJ05,Yang+07}
used an iterative method to select groups, computing a
density enhancement to assign galaxies to groups, starting with seed
groups obtained from the FoF implementation of \citet{Eke+04}. \citet{MM12}
also used an iterative method that can be compared to a FoF on dark
matter halos, starting with the assumption that all galaxies are their own
halo (i.e.\ all groups have a single galaxy in the initial step).
\citet{DominguezRomero12} also started with galaxies being alone in their
groups, and adapted the \citet{Yang+07} algorithm by not directly assigning
galaxies to groups, but computing instead probabilities that galaxies are in
a given group, allowing galaxies during the iterative process to ``move''
between groups; but they assigned galaxies to groups after the convergence
of their iterative method. Finally, in \citet{DM14b}, we have developed
MAGGIE (Models and Algorithm for Galaxy Groups, Interlopers and
Environment), another Bayesian and fully probabilistic grouping algorithm,
which does not make use of the FoF technique.

Nevertheless, the FoF algorithm is still widely used, because the
aforementioned Bayesian algorithms are not publicly available and are quite
difficult to code on one's own. Moreover, the FoF algorithm has the
advantage of providing unique group catalogs (in some other methods, the
group catalog depends on the galaxy one starts with), and makes no
assumption on the properties of groups (i.e.\ number density profile or
three-dimensional shape).

Many catalogs of galaxy groups have been constructed from redshift space
catalogs,\footnote{\cite{TG76} applied a grouping algorithm in projected
space that turned out to be a Friends-of-Friends algorithm.} using FoF
algorithms (\citealp*{HG82,NW87,RGH89,TrasartiBattistoni98,MZ02};
\citealp{Eke+04,Berlind+06,Tago+10,Robotham+11,Tempel+14}). Because of the redshift distortions,
the physical linking lengths are chosen to be of order of 10 times
longer for the LOS separations than for the POS ones. Moreover, for
flux-limited galaxy catalogs, the physical linking lengths are scaled with the mean
three-dimensional separation between neighboring galaxies, $s \simeq n
^{-1/3}$, where $n$ is the mean number density of galaxies in the Universe
at a given redshift \citep{HG82}. In other words, the FoF algorithm involves
two dimensionless linking lengths (hereafter LLs):
\begin{eqnarray}
b_\perp &=& {{\rm Max} (S_\perp)\over s} \ ,
\label{bperpdef}
\\
b_\parallel &=& {{\rm Max} (S_\parallel) \over s} \ ,
\label{bpardef}
\end{eqnarray}
where $S_\perp$ and $S_\parallel$ are the POS
and LOS nearest neighbor separations, respectively.

Starting with~\cite{NW87},  nearly all FoF group analyses
on redshift space catalogs were accompanied with tests on mock galaxy
catalogs derived from N-body simulations. However, not all FoF developers
have applied the same tests to calibrate their linking lengths.~\cite{NW87}
were the first to compute the accuracy of group masses, as well as radii and
velocity dispersions, crossing times and mass-to-light ratios.~\cite{RGH89}
were the first to test the recovered group multiplicity function.
\cite{Frederic95a} was the first to measure the galaxy reliability of
extracted groups (comparing the FoFs of \citealp{HG82} and \citealp{NW87}),
as later done by~\cite{MZ02}, who also measured group completeness (against
mergers of true groups) and reliability (against fragmentation of true
groups). \citet{Eke+04} also tested the true group completeness and
fragmentation, as well as the accuracy on group sizes and velocity
dispersions. They also considered a quality criterion that amounts to a
combination of galaxy completeness and reliability. Finally,
\citet{Berlind+06} performed similar tests as \citeauthor{Eke+04}, with
another test combining galaxy completeness and reliability.
\citeauthor{Berlind+06} noted that one cannot simultaneously optimize the
accuracies on group sizes, velocity dispersions and [multiplicity function OR
combined galaxy completeness/reliability].

Unfortunately, none of these studies is fully convincing: many did not
perform the full suite of important tests, which we believe are true group
fragmentation (group reliability) and merging (group completeness),  galaxy
completeness and reliability studied separately, and mass accuracy. Many
have not measured the qualities of their LLs in terms of group parameters
such as estimated mass and richness. Few studies
have \emph{optimized} the LLs:
\cite{Eke+04} separately optimized
\bperp{} and \bpar. 
\cite{Berlind+06} jointly optimized \bperp{} and \bpar on a grid, for groups of 10 or more galaxies, while
\cite{Robotham+11} jointly fit the LLs and their variation with density
contrast and galaxy luminosity for groups of 5 or more galaxies to
optimize for the product of four fairly complex measures of group and galaxy
completeness and reliability. However, there is no strong agreement between the
optimized LLs of \citeauthor{Eke+04}, \citeauthor{Berlind+06}, and
\citeauthor{Robotham+11} (see Table~\ref{groupalgos}).

Moreover, we believe that in this era of large  redshift surveys of
 $>10^5$ galaxies, it makes
little sense to extract groups from flux-limited galaxy samples, for which
most current implementations of the FoF algorithm scale the maximum
separations proportionally to the mean  separation between neighboring field
galaxies, $n^{-1/3}$. Indeed, since the minimum luminosity in flux-limited
samples increases with redshift,  the mean number density of galaxies
decreases with redshift, and thus the mean separation between neighboring galaxies
increases with redshift. Therefore, the standard implementation of the FoF
algorithm leads to groups that become increasingly sparse and with
increasingly higher velocity dispersion with redshift (while their
multiplicity function is preserved).
Alternatively, since the mean neighbor galaxy separation increases with redshift in
flux-limited samples, using a fixed physical linking length leads to lower reliability at low
redshift and lower completeness at higher redshifts.
Moreover, grouping algorithms on
flux-limited samples must evaluate the luminosity incompleteness  as a
function of redshift, which is difficult and imprecise (e.g.,
\citealp{MDNC02,Yang+07}). It is therefore much safer to consider subsamples
that are complete in both distance and galaxy luminosity (as done for FoF
grouping by
\citealp{Berlind+06}, \citealp{Tago+10} and \citealp{Tempel+14}).
Admittedly, one recovers at best of order of one-quarter of the galaxies of
the flux-limited sample, but one then avoids extracting a heterogeneous
sample of groups (see \citealp{Tempel+14})
whose sizes and velocity dispersions stretch with redshift
(when scaling the physical linking lengths with $n^{-1/3}$) or whose
completeness and reliability vary with redshift (when adopting fixed
physical linking lengths).

In the present work, we shall provide the first optimization of group
LLs for doubly complete subsamples of galaxies, for six measures of the quality of the
FoF grouping algorithm: minimal fragmentation and merging of true groups,
maximum completeness and reliability of the galaxies of the extracted
groups, and minimum bias and inefficiency in the recovered group masses.
These tests are performed on a wide grid of over 250 pairs of LLs. We have
applied them to several doubly-complete subsamples of galaxies cut from a
mock flux-limited, SDSS-like,  sample of galaxies, and we analyze our
results in terms of both true and estimated masses of the groups, as well as
of their estimated richness.

We present  the FoF algorithm in Sect.~\ref{sec:fof} and make predictions on
its optimal parameters. In Sect.~\ref{sec:mock}, we describe our mock
real-space galaxy and group catalogs, and explain how we extract our mock
redshift-space group catalogs. We define our tests in Sect.~\ref{sec:tests}.
In Sect.~\ref{sec:results}, we present the results of our tests, comparing
to various grouping methods, and suggest an optimal set of LLs. Finally, we
summarize and discuss our results in Sect.~\ref{sec:discussion}.
\section{Friends-of-Friends algorithm}
\label{sec:fof}
\subsection{Predicted linking lengths and galaxy reliability}
\label{sec:fofpred}
One can relate the choice of $b_\perp$ to the minimum galaxy overdensity (in
number) of the groups with
\begin{equation}
{\delta n\over n} = {3\over 4 \pi b_\perp^3} -1 \ ,
\label{dnovernfromb}
\end{equation}
(from \citealp{HG82}). Hence, if galaxies are unbiased tracers of mass, i.e.
$\delta n/n = \Delta/\Omega_{\rm m}$, where $\Omega_{\rm m}$ is the
cosmological density parameter, then equation~(\ref{dnovernfromb}) easily
leads to
\begin{equation}
b_\perp = {\left ({3/(4\pi)\over \Delta/\Omega_{\rm m}+1}\right)}^{1/3}\,.
\label{bperpfromDelta}
\end{equation}

According to equation~(\ref{bperpfromDelta}), if one desires to have
virialized groups of overdensity (relative to critical) $\Delta=200$,
one requires $b_\perp\simeq 0.07$ (for $0.24 <
\Omega_{\rm m} < 0.35$). On the other hand, given $\Omega_{\rm m}=0.279$ or
0.317, respectively obtained with the 9th-year release of the Wilkinson
Microwave Anisotropy Probe \citep{Bennett+13} and the Planck mission
\citep{PlanckXVI}, one deduces $\delta n/n=352$ and 326 from
\citeauthor{BN98}'s (\citeyear{BN98})
approximation for $\Delta$ at the virial radius~, leading to $b_\perp \simeq 0.09$ in both
cases, according to equation~(\ref{dnovernfromb}).

One can also estimate the ratio of LOS to transverse LLs, as the ratio of LOS
to POS group sizes caused by redshift distortions: if the LOS velocities span
$\pm \kappa$ group velocity dispersions, the inferred LOS spread of distances
in redshift space will be $\pm \eta \kappa\, v_{200} / H_0 = \pm \eta \kappa
\sqrt{\Delta/2}\,r_{200}$ (see \citealp{MBM10}),
where  $\eta = \sigma_v/v_{\rm v} \simeq 0.65$ for an NFW model with
realistic concentration and velocity anisotropy \citep{MBB13}, and where we
used equation~(\ref{dnovernfromb}).
Therefore,
%
\begin{eqnarray}
{b_\parallel\over b_\perp} &=&
\eta \,\kappa\, \sqrt{\Delta \over
    2}
\label{LLratiofromDelta}\\
&=&
\eta \,\kappa\, \sqrt{{\Omega_{\rm m}\over
    2}\,\left({\delta n\over n}\right)} \;.
\label{bparfromkappa}
\end{eqnarray}
Combining
equations~(\ref{bperpfromDelta}) and (\ref{LLratiofromDelta}), one easily
deduces
\begin{equation}
\kappa = \sqrt{8\pi\over 3}\,\eta^{-1}\,\Omega_{\rm m}^{-1/2}\,
\sqrt{b_\perp}\,b_\parallel \,.
\label{kappafromlls}
\end{equation}

For example, according to equation~(\ref{LLratiofromDelta}), probing galaxies
along the LOS to $\pm1.65\,\sigma_v$ (encompassing 95\% of the galaxies for
Maxwellian LOS velocity distributions), for $\Delta = 200$, leads to
$b_\parallel/b_\perp=11$, hence with $b_\perp=0.07$, one finds
$b_\parallel=0.7$ (the values are rounded off).

These theoretical LLs assume that groups are spherical and that all but one
galaxy is in the center. In fact,  galaxies are distributed in a more
continuous fashion (especially in rich groups and clusters).
One can more accurately estimate the value of the transverse LL by
writing
\begin{eqnarray}
b_\perp &=& {{\rm Max}(S_\perp)\over  n^{-1/3}}
\ ,
  \nonumber \\
&=& {{\rm Max}(S_\perp)\over r_{{\rm vir}}}\,
{r_{{\rm vir}}\over  n_{\rm vir}^{-1/3}}\,
{\left(1+{\delta n\over n}\right)}^{-1/3}
 \ , \nonumber \\
&=& {\left({3/(4\pi)\over \Delta/\Omega_{\rm m}+1}\right)}^{1/3}\,
{{\rm Max}(S_\perp)\over
    r_{{\rm vir}}}\,N_{\rm vir}^{1/3} \ ,
\label{bperppred2}
\end{eqnarray}
where one recognizes the previous estimate of $b_\perp$
(eq.~[\ref{bperpfromDelta}]) in the first term of the right-hand side of
equation~(\ref{bperppred2}).

We
estimated the value of the second term of the right-hand side of
equation~(\ref{bperppred2}) by running Monte-Carlo simulations of
cylindrical groups of unit virial radius with surface density profiles
obeying the (projected) NFW model of scale radius of 0.2 (i.e.\
concentration 5). With $10\,000$ realizations each for $N=2, 4, 8, 16, 32$
and 64 galaxies within  the maximum projected radius allowed for
the galaxies in the simulated groups, $R_{\max}=r_{200} =1$,
we found that the  95th percentile for the maximum {--} for all galaxies
of the group {--} distance to the nearest neighbor is ${\rm Max}(S_\perp)
\simeq 1.48\,N^{-0.25}$ in units of the virial radius.
Inserting
this value of ${\rm Max}(S_\perp)/r_{{\rm vir}}$ into
equation~(\ref{bperppred2}), with $\Delta=200$ and $\Omega_{\rm m}=0.25$, we
predict that to obtain a completeness of 0.95, we require
\begin{equation}
b_\perp \simeq 0.09\,N^{0.08} \ ,
\label{bperppred3}
\end{equation}
where we took into account that, for our adopted NFW model, the ratio of the
number of galaxies within the virial sphere to that within the virial
cylinder is $N_{\rm vir}/N\simeq 0.80$. Equation~(\ref{bperppred3}) predicts
$b_\perp=0.10$ for $N=4$ and $b_\perp=0.12$ for $N=40$, i.e.
$b_\parallel=1.1$ for $N=4$ and $b_\parallel=1.3$ for $N=40$, given
$b_\parallel/b_\perp=11$ found above. In other words,
equation~(\ref{dnovernfromb}) underestimates $\delta n/n$ by a factor ${\rm
Max}(S_\perp)/r_{{\rm vir}}\,N_{\rm vir}^{1/3} \simeq 1.4\,N^{0.08}$, i.e.\
by 1.5 for $N=4$ and 1.8 for $N=40$. The slight increase of $b_\perp$ with
richness suggests that fixing $b_\perp$ will lead to the fragmentation of
rich groups.

Adopting instead the virial $\delta n/n = \Delta/\Omega_{\rm m} = 326$
(Planck, see above) would
lead to $b_\perp = 0.14$ for $N=4$ and $b_\perp=0.17$ for $N=40$.
Since, at constant $\Delta$, $b_\perp \propto \Omega_{\rm m}^{1/3}$
(eq.~[\ref{bperpfromDelta}]),
moving from $\Omega_{\rm m}=0.25$ to
 $\Omega_{\rm m}=0.3$ (a compromise between
WMAP and Planck), keeping $\Delta=200$,  yields $b_\perp=0.11$ ($N=4$) or 0.13 ($N=40$). 
According to equation~(\ref{LLratiofromDelta}), $b_\parallel/b_\perp$ does
not vary with $\Omega_{\rm m}$ at fixed $\Delta$, hence we now obtain
$b_\parallel = 1.3$.

Had we taken a maximum projected  radius that is much smaller than
$r_{200}$, we would obtain a much smaller value for $b_\perp$. Indeed, our
Monte-Carlo simulations indicate that with $R_{\max}$ and scale radius both
equal to $0.2\,r_{200}$, we find ${\rm Max}(S_\perp)
\simeq 1.85\,N^{-0.33}$ in units of $R_{\max}$, hence ${\rm
Max}(S_\perp)/r_{200} \simeq 0.37\,N^{-0.33}$. Inserting this ratio into
equation~(\ref{bperppred2}), we now obtain $b_\perp = 0.023$, independent of
$N$. Thus, to first 
order, $b_\perp$ scales with $R_{\max}/r_{200}$. Turning the argument
around, a low $b_\perp$ leads to selecting galaxies in groups with projected
radii limited to a small fraction of the virial radius.

We can also predict the reliability of the galaxy membership in groups, as
follows. The expected number of interlopers from the extracted group out to a LOS
distance of $\pm b_\parallel n^{-1/3}$ is
\begin{equation}
N_{\rm int} \approx 2\,{N\over 200} \,{b_\parallel\over b_\perp} \ ,
\label{Nilop}
\end{equation}
where we simply stretched the group by a factor of $b_\parallel/b_\perp$
along the LOS, and where $N$ is the number of galaxies in the real space group.
For $b_\parallel/b_\perp=11$, equation~(\ref{Nilop}) yields $N_{\rm int}=
0.44$ for $N=4$ and $N_{\rm int}=4$ for $N=40$. Thus, the fraction of
interlopers should roughly be independent of the richness hence mass of the
real space group.
For $b_\perp \simeq 0.1$, corresponding to groups with overdensity 200
relative to critical sampled at 95\% completeness,
and sampling the LOS with 95\% completeness (leading to
$b_\parallel/b_\perp=11$), one then expects $N_{\rm int}/N = 0.11$.
One then infers a galaxy reliability of $R=(N/N_{\rm
int})/[1+(N/N_{\rm int})] = 90\%$.

Equation~(\ref{Nilop}) assumes that the Universe is made of spherical groups
that are truncated at their virial radii. In fact, galaxy clustering brings
galaxies close to groups, in a fashion that the radial number density
profile pursues a gradual decrease beyond the virial radius. For NFW models
of concentration of 5,
the projected number of galaxies within the virial radius is 
$1/0.80 = 1.25$ times the number within the virial sphere. Hence the numbers of
interlopers to the virial sphere should satisfy $N_{\rm int}/N=0.25$. Then,
one expects a reliability of $R=(N/N_{\rm int})/[1+(N/N_{\rm int})] = 80\%$.
\subsection{Previous implementations}
Table~\ref{groupalgos} lists the dimensionless LLs for the different group
FoF analyses.
\begin{table}
\caption{Friends-of-Friends linking lengths and physical parameters}
\begin{center}
\setlength{\tabcolsep}{2.5pt}
\begin{tabular}{llllcrl}
\toprule%
\toprule%
Authors & sample & \multicolumn{1}{c}{$b_\perp$} & \multicolumn{1}{c}{$b_\parallel$} &
\multicolumn{1}{c}{$b_\parallel/b_\perp$} & $\delta n/
n$ & \multicolumn{1}{c}{$\kappa$} \\
\toprule%
Huchra \& Geller 82     & CfA     & 0.23  & 1.34  & \ \ \ \ 6.3  & 20    & 5.7 \\
Ramella et al. 89       & CfA2    & 0.14  & 1.9   & 13   & 80    & 5.8 \\
Trasarti-Battistoni 98  & PPS2    & 0.13  & 1.7   & 13   & 108   & 4.9\\
Merchan \& Zand'z 02    & 2dFGRS  & 0.14  & 1.4   & 10   & 80    & 4.4 \\
Eke et al. 04           & 2dFGRS  & 0.13  & 1.43  & 11   & 178   & 3.9\\
Berlind et al. 06       & SDSS    & 0.14  & 0.75  & \ \ \ \ 5.4  & 86    & 2.3 \\
Tago et al. 10          & SDSS    & 0.075  & 0.75  & 10   & 565  & 1.7\\
Robotham et al. 11      & GAMA    & 0.060  & 1.08  & 18   & 1100  & 2.2\\
Tempel et al. 14 ($M_r$$<$$-19$)        & SDSS    & 0.11  & 1.1 & 10 & 178 & 3.0 \\
Tempel et al. 14 ($M_r$$<$$-21$)        & SDSS    & 0.066  & 0.67 & 10 & 830 & 1.4
\\
\bottomrule%
\end{tabular}
\end{center}
\parbox{\hsize}{%
Notes: The (normalized) linking lengths of~\cite{HG82},~\cite{RGH89},
    and~\cite{TrasartiBattistoni98} are derived (using eqs.~[\ref{bperpdef}]
    and [\ref{bpardef}]) from their physical linking lengths at the fiducial
    distance and from the mean density at that distance, as derived by
    integrating the respective luminosity functions given by these authors.
    The linking lengths of~\cite{MZ02} are estimated directly from the
    overdensity $\delta n/n$ given by these authors (using
    eq.~[\ref{dnovernfromb}]), those of~\cite{Tago+10} are found from the
    densities deduced from the numbers of galaxies counted by these authors (again with
    eq.~[\ref{bperpdef}] and [\ref{bpardef}]).~\cite{Eke+04} provide
    $b_\perp$ and $b_\parallel/b_\perp$, while \cite{Berlind+06} and
    \cite{Tempel+14} provide
    $b_\perp$ and $b_\parallel$. When not provided by the authors, the
    overdensity $\delta n/n$ is obtained through
    equation~(\ref{dnovernfromb}), and should be multiplied by 1.5 for a more
    accurate estimation (see text). Finally, the number of group velocity
    dispersions along the LOS, $\kappa$ is obtained with
    equation~(\ref{kappafromlls}) assuming $\Omega_{\rm m}=0.3$.
}
\label{groupalgos}
\end{table}
The values of $\delta n/n$ and $\kappa$ of different FoF analyses, inferred
from their LLs according to equations~(\ref{dnovernfromb}) and
(\ref{bparfromkappa}), are listed in Table~\ref{groupalgos}. One sees that 5
of the 7 previous studies advocate $b_\perp = 0.13$ or 0.14, and two
(\citealp{Eke+04} and \citealp{Tempel+14} for $M_r < -19$) have pairs
of LLs close
to our
predicted values of $(b_\perp,b_\parallel)\approx (0.11,1.3)$. The two
greatest outliers are \cite{HG82}, whose transverse linking length appears too large
and~\cite{Robotham+11}, both of whose LLs appear too small. We will check
these conclusions in Sects.~\ref{sec:results} and~\ref{sec:discussion} using
our analysis of mock galaxy and group catalogs.

\subsection{Practical implementation of the FoF algorithm}
\label{sec:algo}
There are two issues that need to be optimally handled when writing an FoF
algorithm: rapidly extracting the separations in redshift space and properly
estimating the mean density.

We followed the~\cite{HG82} algorithm, used in most FoF implementations.
\citeauthor{HG82} write that two galaxies with redshifts $z_i$ and $z_j$ and
an angular separation in $\theta_{ij}$ are linked using criteria that amount
to
\begin{eqnarray}\label{eq:crit1}
    \left({c \over H_0}\right)\left(z_i + z_j\right)\sin\left({\theta_{ij}\over2}\right)
    & \leq & b_\bot \, n^{-1/3} \ , \\
\label{eq:crit2}
    \left({c \over H_0}\right)\left|z_i-z_j\right| &\leq & b_\parallel
    \, n^{-1/3}
\,.
\end{eqnarray}

We generalized\footnote{The \emph{comoving distance}, $d_{\rm comov}(z) =
    c\,\int dz/H(z)$, in equation~(\ref{thetamax}) should really be the
    \emph{proper motion distance} $d_{\rm pm} (z) = d_{\rm lum}(z)/(1+z) =
    (1+z) \,d_{\rm ang}(z)$, but for flat cosmologies, $d_{\rm pm}(z) =
d_{\rm comov}(z)$.} equations~(\ref{eq:crit1}) and (\ref{eq:crit2})
to\footnote{Equation~(\ref{thetamax}) is similar to the relation used
    by~\cite{Zandivarez+14}, with the exception of a minor difference in
projected sizes given angle.}
\begin{eqnarray}
{d_{\rm comov}(z_1) + d_{\rm comov}(z_2)\over 2}\,\theta
&\leq&  b_\perp\,  n^{-1/3} \ ,
\label{thetamax}
\\
|d_{\rm comov}(z_1)-d_{\rm comov}(z_2)| &\leq&
b_\parallel\, n^{-1/3} \,.
\label{deltadmax}
\end{eqnarray}
Thus,~\cite{HG82} and~\cite{Berlind+06} neglected cosmological effects. For
our deepest mock SDSS catalog, at $z = z_{ \max } = 0.125$ (Catalog 6, see
Table~\ref{table:samples} below), $d_{\rm comov}/(cz/H_0) = 0.97$. So, the
formula $d=c z/H_0$ leads to slightly too large distances, hence to
slightly too strict choices of angles and differences in redshifts.

One could argue that, since groups are virialized, one ought to use the
cosmological \emph{angular distance}, $d_{\rm ang}(z) = d_{\rm
comov}(z)/(1+z)$ for the distances with which one computes the physical
transverse separation in terms of the angular separation. But one should
then also compress the line-of-sight distances accordingly, and we are not
aware of any work doing such a compression. Hence, we chose to stick with
equations~(\ref{thetamax}) and (\ref{deltadmax}).

Since we are working with samples that are complete in luminosity, and since
they are shallow enough that evolutionary effects are small, observers can
estimate the mean number density of galaxies directly from the data. 

Finally, for each galaxy, we computed the maximal angular distance to define
the region in which potential neighbors could be found for the given
transverse linking length. With the  celestial sphere grid that we have
constructed (see Appendix~\ref{app:grid}), we searched for galaxies obeying
the criterion of equation~(\ref{thetamax}), and then searched for galaxies meeting
equation~(\ref{deltadmax}). The linked galaxies were then placed in a tree
structure according to the Union-Find method \citep{Tarjan84}. Once all
galaxies were analyzed, we compressed the trees constructed with linked
galaxies by replacing, in each group, the links of links with links to a
single galaxy, giving us the identity of the group to which galaxies belong
to. This implementation allows for a fast computation of galaxy groups for
large samples of galaxies.
\section{Mock catalogs}
\label{sec:mock}
We wish to check if galaxy groups extracted with FoF algorithms are
optimally selected.  So our goal is to compare the \emph{extracted groups}
(EGs) in redshift space with the \emph{true groups} (TGs) in real space.
Since real space information is not directly accessible,  we need to
simulate it. The best way is to use mock galaxy catalogs constructed from
the outputs of realistic galaxy simulations. These should include real space
galaxy positions, comoving velocities, stellar masses and $r$-band
luminosities, and the galaxies should be assembled in (real-space) groups
with realistic density profiles and obeying the observed scaling relations.
We then need to construct a redshift space catalog of groups from the real
space catalog of galaxies and groups.
\subsection{Construction of mock real-space galaxy and group catalogs}
There are two basic methods to build a mock catalog of galaxies in real space.
\begin{enumerate}
        \item In the Halo Occupation Distribution method \citep{MS02,BW02},
            the number of galaxies per halo is drawn from a probability
            distribution function that depends on the halo mass, or better,
            the galaxy luminosities or stellar masses are drawn from
            conditional luminosity (stellar mass) functions that depend on
            halo mass \citep{YMvdB03}.  The galaxy distribution is assumed to be spherically
            symmetric, and  follows that of the dark matter particles in the
            halos of $\Lambda$CDM cosmological simulations (e.g., NFW), the
            velocities are drawn from Maxwellian distributions (see
            \citealp{Beraldo+14} for the limitations of this assumption),
            with radial and tangential  velocity dispersions derived from
            the Jeans equation of local dynamical  equilibrium, assuming
            some form for the radial variation of the velocity anisotropy.

        \item In  Semi-Analytical Models (SAMs, e.g.,
            \citealp{RQPR97,KCDW99}), galaxy properties (in particular
            stellar mass and $r$-band luminosity) are painted on the halos
            and subhalos of cosmological $N$ body simulations across cosmic
            time, following well-defined recipes for star formation and
            galaxy feedback.  This procedure produces galaxies that follow
            relatively well the observed properties and scaling relations.
\end{enumerate}
We have chosen the second approach, because the recent SAM by~\cite{Guo+11},
run on the Millennium-II simulation \citep{BoylanKolchin+09} fits well the
$z$=0 observations (as shown by \citeauthor{Guo+11}). The Millennium-II
simulation  involved $2160^3$ particles in a box of comoving size 137 Mpc,
running with cosmological parameters $\Omega_{\rm m}=0.25$,
$\Omega_\Lambda=0.75$, $h=0.73$, and $\sigma_8=0.9$. The particle mass was
thus $9.5\times 10^6 M_\odot$.

We extracted the SAM output of~\cite{Guo+11} from the Guo2010a database on
the German Astrophysical Virtual Observatory
website.\footnote{http://gavo.mpa-garching.mpg.de/Millennium/Help, see
\cite{Lemson06}} The real-space TGs were extracted by \citeauthor{Guo+11}
using the FoF technique applied to the particle data, with over $10^5$
particles for groups of mass $>10^{12} \rm M_\odot$. The database includes
the mass within the sphere of radius $r_{200}$, where the mean mass density
is $\Delta=200$ times the critical density of the Universe, centered on the
particle in Millennium-II simulation, within the largest subhalo, with the
most negative gravitational potential \citep{BoylanKolchin+09}. We slightly
modified the membership of the TGs by considering only the galaxies within
$r_{200}$.\footnote{We kept the galaxies outside the sphere of radius
$r_{200}$ as possible interlopers.}

\subsection{Construction of mock redshift-space group catalogs}
\label{sec:mockzspace}
\begin{figure}
    \centering
    \includegraphics[width=0.7\linewidth]{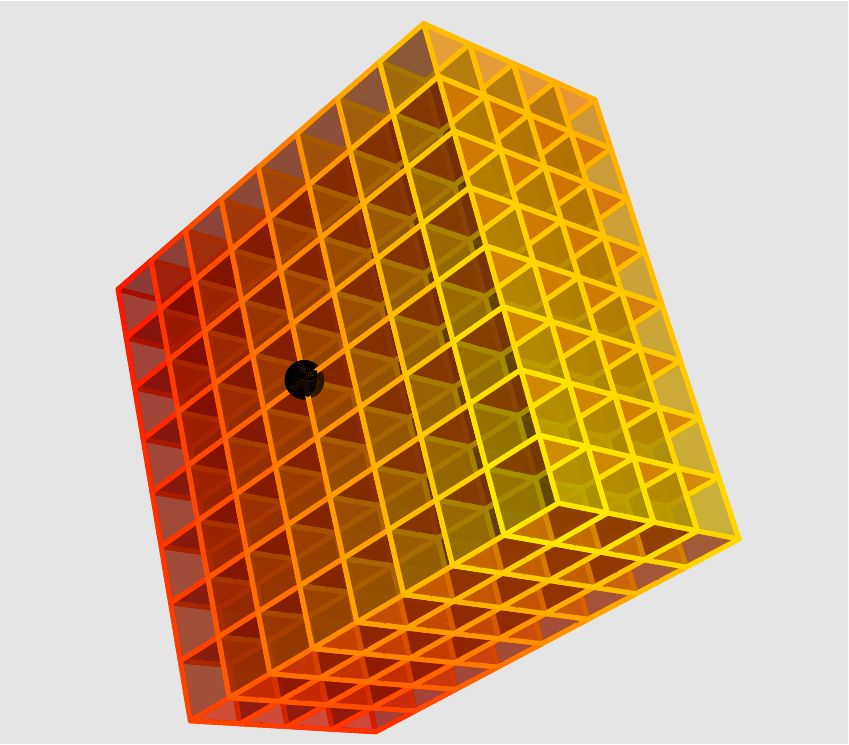}
    \caption{A representation of the full mock galaxy catalog. Each small box is
        a  137 Mpc long cube. The observer is at the \emph{large black
        point} in the middle of one of the square sides of the superbox.
    }
\label{fig:mock}
\end{figure}

We now describe the construction of the mock SDSS redshift space galaxy
catalog.
We first note that our simulation box is not large enough to produce a deep
enough redshift-space group catalog.  Indeed, the simulation box size limits
the view to $z=0.034$ from one corner to the next, or to $z=0.058$ along the
longest diagonal.  We therefore replicated the simulation boxes along the
three cartesian coordinates to reach our desired maximum radius, thus
creating a \emph{superbox}.
Moreover, since the SDSS survey is wider than $\pi/2$ sr (our mock SDSS has
a solid angle of 2.2 sr), we could not place the observer at the corner of
the superbox.  Instead, we placed the observer at the middle of one of the
sides of the superbox.  Then, the size of the superbox must be double the
proper distance of 505 Mpc to the highest redshift that we wish to sample,
$z=0.126$ (Table~\ref{table:samples}), along two directions, and 505 Mpc in
the third (LOS) direction.  One thus requires replicating the
simulation cubes for a total of  $4\times 8\times 8=256$ simulation cubes in
our superbox
(see Figure~\ref{fig:mock}).
Because the redshifts are small, we only made use of the $z$=0 simulation
box, thus neglecting the small late evolution of group
properties.

As pointed by \citet{Blaizot+05}, this procedure of replicating can cause
structures to appear periodically along the LOS\@. To avoid this feature,
we followed \citeauthor{Blaizot+05}, applying  random transformations on the
boxes: $\pm\pi/2$ rotations around the 3 cartesian axes, random periodic
translations and random mirror reflections along a given axis.  These
transformations were applied to the phase space coordinates of galaxies.

We derived the coordinates of the galaxies on the sky, using standard
formulae of spherical trigonometry. Absolute magnitudes were converted to
apparent magnitudes, and the flux limit of the primary spectroscopic sample
of the SDSS, $r < 17.77$, was applied. We assumed here that the observer
knows how to correct his sample for Galactic and internal extinction, as
well as k-correction, hence no backwards corrections were applied to our
mock galaxies.

From this flux-limited sample, we constructed subsamples that are doubly
complete in distance and luminosity.\footnote{This step of flux-limited
sample can be avoided, but serves to show that our doubly complete
subsamples are taken from the same parent sample.}

We added the Hubble flow corresponding to the value of the Hubble constant used
in the Millennium-II ($h=0.73$).
For this, we did not immediately compute LOS velocities. Instead,
we derived the galaxy redshifts, $z$, by
first solving for the redshift $z_{\cos}$ that a galaxy would have with zero
peculiar velocity:
\begin{equation}
d_{\rm comov}(z_{\cos}) = d \ , \\
\end{equation}
(where $d$ is the Euclidean distance to the observer in the superbox)
and then by determining the redshift given the galaxy's LOS peculiar velocity,
$v_{\rm p}^{\rm LOS} = \bmath{v_{\rm p}} \cdot \bmath{d}/d$,
with \citep{HN79}
\begin{equation}
1+z = \sqrt{1+v_{\rm p}^{\rm LOS}/c\over 1-v_{\rm p}^{\rm LOS}/c}\,\left(1+z_{\cos}\right)
\,.
\label{zfromvp}
\end{equation}

We did not consider the SDSS limit on surface brightness, as it only affects
a small fraction of the galaxies and surface brightness is not very well
defined in the outputs of the SAM\@.

EG catalogs constructed as described above have 2 sets of unavoidable
artefacts:
%
1.
TGs that lie close the edges of the simulation box can be split during
    the process of random rotation, reflection and translation of the boxes.
2. Since the SDSS survey is not all-sky, TGs can be cut by the edges of
    the survey.
We therefore first flagged the groups in real space that were split during the
transformations (translations and rotations) of the simulation box.
We neglected holes in our
survey mask caused by spectroscopic incompleteness, bright stars, camera
problems, etc., for technical simplicity.  For example, the spectroscopic
incompleteness is more present on dense regions on the celestial sphere
because of more frequent fiber collisions.  Simulating this would require
the calibration of incompleteness as a function of density in the SDSS sky
and then apply it to our mock. This is complex and may not be accurate.
Also,~\cite{Berlind+06} found that fiber collisions only caused a small
decrease (0.06 dex) of the group multiplicity function. Moreover, our goal
is to test the FoF technique in a perfect situation, where all observational
errors are neglected.

\begin{figure*}
    \centering%
    \subfloat[Projected space galaxy mock catalog]
    {%
        \includegraphics[width=0.49\linewidth]{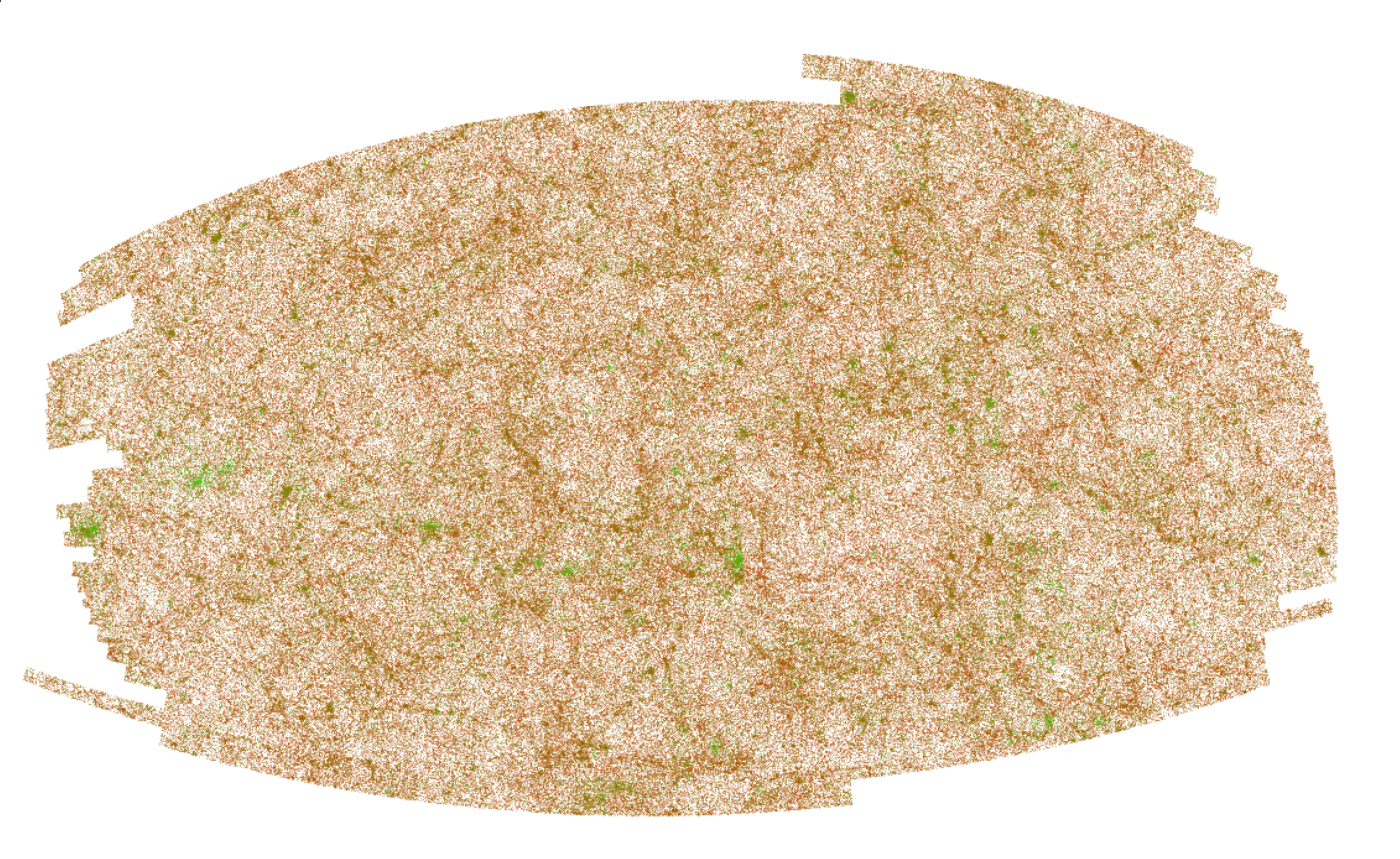}%
    }%
    \subfloat[Redshift space galaxy mock catalog]
    {%
        \includegraphics[width=0.49\linewidth]{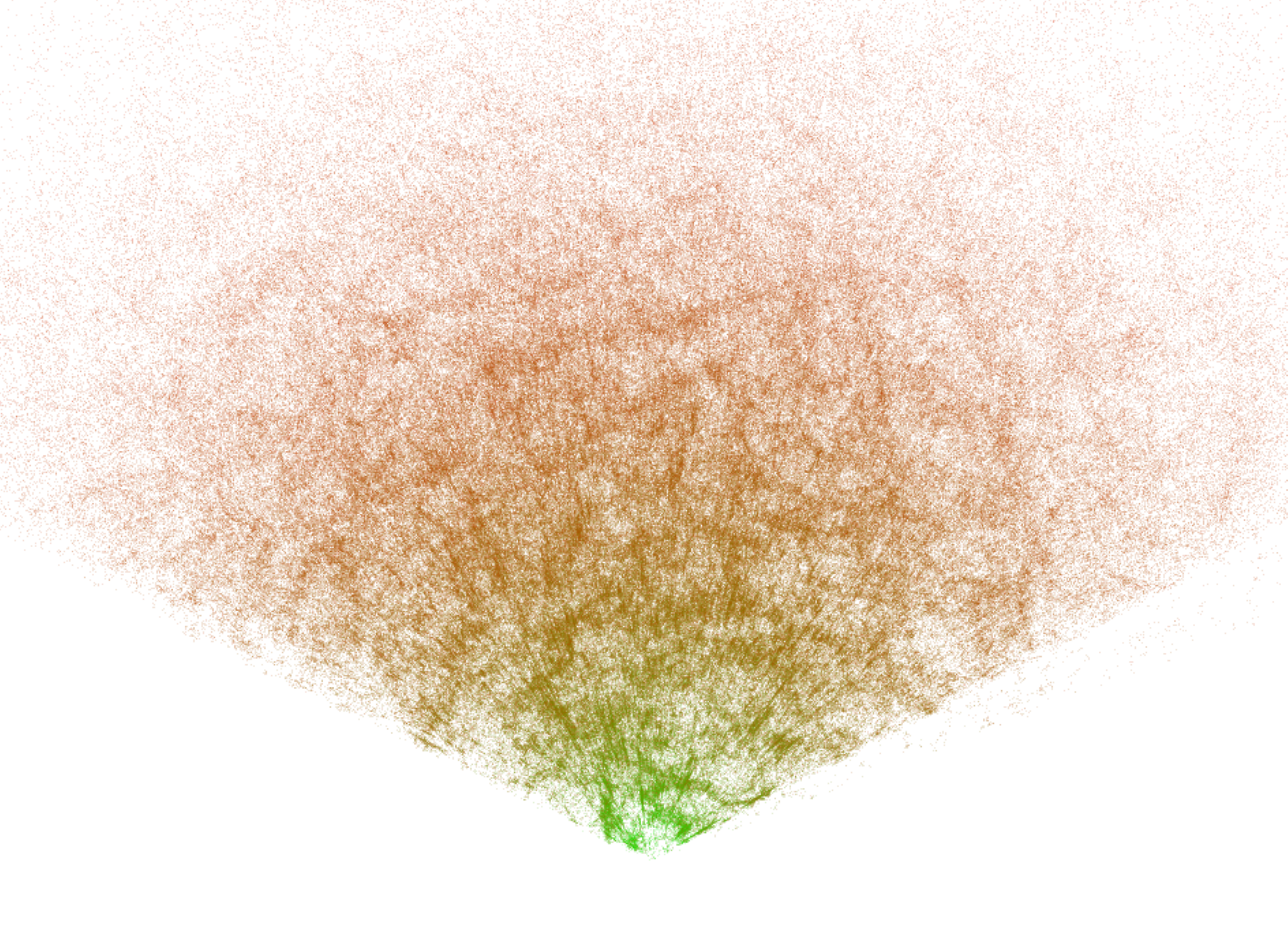}%
    }%
\caption{Views of our initial, flux-limited, 2.2 sr SDSS mock galaxy catalog,
  projection on the
  celestial sphere (\emph{left}) and 3D cone (\emph{right}, not sliced).
The colors provide the absolute  $r$-band magnitude (\emph{green} for low luminosity).
}
\label{fig:mask}
\end{figure*}

The resulting mock flux-limited catalog, shown in Figure~\ref{fig:mask}, contains
823$\,$497 galaxies.

%
%
%
%
%
\subsection{Samples}

Finally, we extracted several subsamples of galaxies and groups from our
flux-limited sample, using half-integer values for the faintest absolute
magnitude. We also adopted a minimum redshift of $z=0.01$. Otherwise, at
lower redshifts,  peculiar
motions of galaxies are non-negligible contributors to their redshifts, and thus
contaminate the distances required to estimate the galaxy luminosities
and stellar masses.

\begin{table}
    \caption{Doubly complete mock galaxy subsamples}
\begin{center}
\setlength{\tabcolsep}{3pt}
\begin{tabular}{lccccccc}
\toprule
\toprule
ID & $M_r^{\max}$ & $L_r^{\min}/L*$ & $z_{ \max }$ & Number & $n$
& $n^{-1/3}$ & Fraction\\
          &            &     &                 &        & ($\rm  Mpc^{-3}$)
& (Mpc) & split\\
\toprule
1 & --18.5 & 0.09 & 0.042 & \ \,47158 & 0.0125 & 4.32 & 5.3\%\\
2 & --19.0 & 0.14 & 0.053 & \ \,72510 & 0.0099 & 4.66 & 6.1\%\\
3 & --19.5 & 0.22 & 0.066 & 112629    & 0.0078 & 5.05 & 6.6\%\\
4 & --20.0 & 0.36 & 0.082 & 166899    & 0.0058 & 5.56 & 7.4\%\\
5 & --20.5 & 0.56 & 0.102 & 213546    & 0.0040 & 6.29 & 8.6\%\\
6 & --21.0 & 0.90 & 0.126 & 245821    & 0.0025 & 7.40 & 9.9\%\\
\bottomrule
\end{tabular}
\end{center}
\parbox{\hsize}{Notes: Columns are: sample, maximum $r$-band absolute
  magnitude, minimum luminosity in units of $L*$
(adopting $M*=-20.44 + 5\,\log h$ in the SDSS $r$ band from \citealp{Blanton+03}),
maximum redshift,
sample size,
mean density $n$, proxy for the mean separation to the closest neighbor,
$n^{-1/3}$, and the percentage of true groups that are flagged because they are
split during the simulation box transformations.
The minimum redshift of each subsample is $z=0.01$.
}
\label{table:samples}
\end{table}

Our adopted doubly-complete galaxy subsamples are listed
in Table~\ref{table:samples}.
Here, the mean density of each subsample is constant within, contrary to the
flux-limited case. 
Subsample 1 spans deepest down the luminosity function to $0.09\,L*$, but has
5 times fewer galaxies than the two most distant samples. However, by only
selecting galaxies more luminous than $0.9\,L*$, subsample 6 is limited to
somewhat rare giant galaxies.

\section{Testing methods}
\label{sec:tests}
We tested the FoF algorithm by running it on our mock redshift-space, doubly
complete subsamples of galaxies, for a set of $16\times16$
geometrically-spaced pairs of LLs. By directly comparing the properties of
our EGs extracted in redshift space with their ``parent'' TGs in  real
space, we could assess the performance of the FoF in recovering the real
space information from the projected phase space observations. Note that TGs
can have as little as one single member galaxy. Also, galaxies in redshift
space with no linked galaxies can be considered as EGs with one single
galaxy.
\subsection{Linking real space and projected redshift space}
There are several ways to link the EGs and TGs. We followed~\cite{Yang+07},
by linking the EG to the TG that contains the EG's most massive galaxy
(MMG), and conversely linking the TG to the EG that contains the TG's MMG\@.
With this definition for linking, we could easily associate FoF groups
to real groups.
\begin{figure}
    \centering
    \includegraphics[width=0.7\hsize]{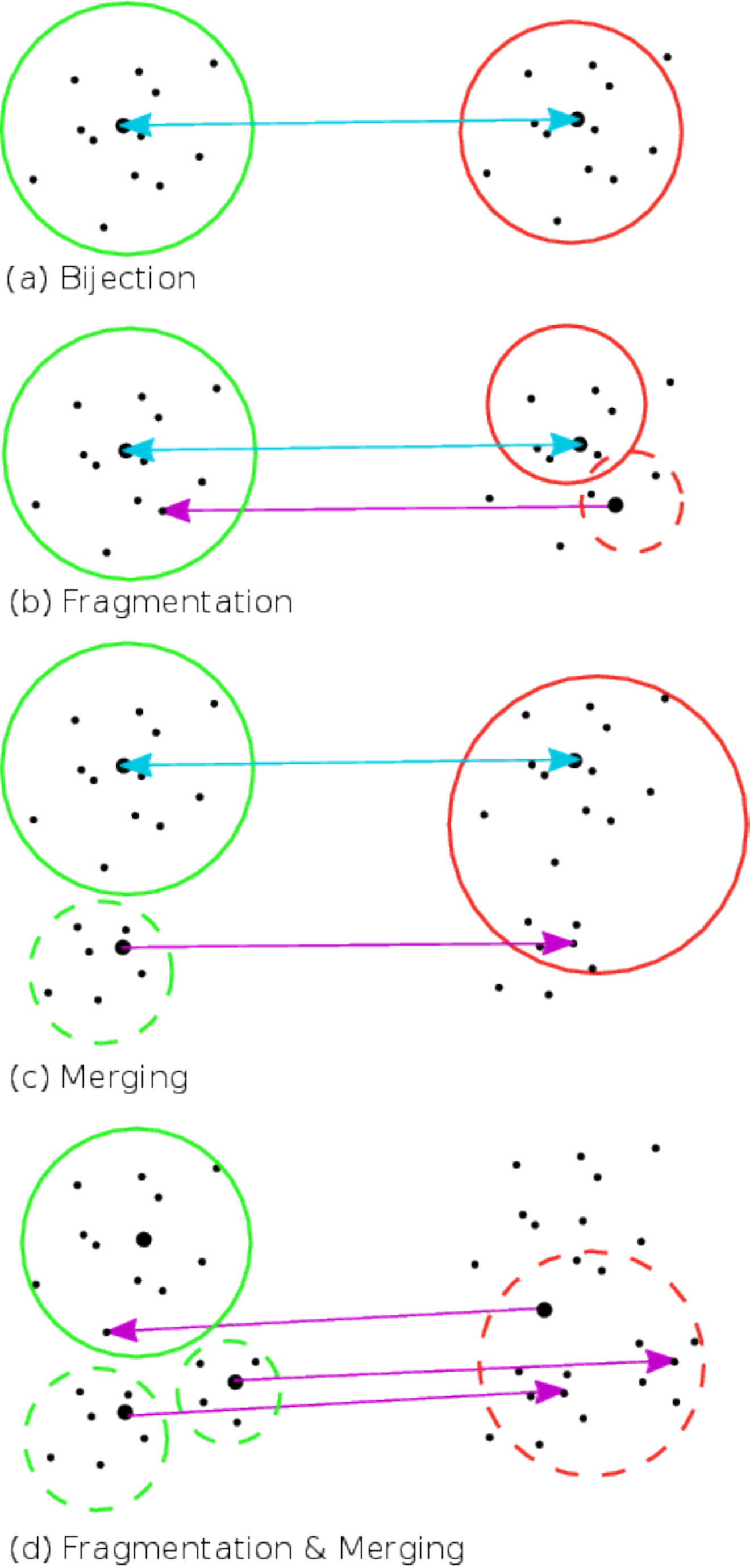}%
    \caption{Schematic links between true groups (\emph{green circles})
    and FoF-extracted groups, (\emph{red circles}),
    each with their respective
    most massive galaxy (\emph{black dots}). 
The \emph{solid circles } represent primary true and FoF groups, while the
 \emph{dashed circles} respectively correspond to
 secondary true groups and FoF fragments.
    The \emph{cyan double arrows}
each indicate the one-to-one
    correspondence between the most massive galaxy in the true and extracted
    groups.
    The \emph{purple rightwards-pointing  arrows}
correspond to the most
    massive galaxy of a true group ending up as a galaxy that is not the
    most massive of its extracted group.
    The \emph{purple leftwards-pointing arrows}
 represent the cases where
    the most massive galaxy of an extracted group is not the most massive of
    its parent true group.
\label{fig:CRdef}}
\end{figure}
\subsection{Global tests}
Our definition of the link between EGs and TGs allowed us to search for
cases where there is no one-to-one correspondence between the groups in real and
redshift space: a TG can suffer from  \emph{fragmentation} into several EGs,
while an EG can be built from the \emph{merging} of several TGs.

Figure~\ref{fig:CRdef} illustrates different cases (following an analogous
figure in \citealp{Knobel+09}). The top panel shows a one-to-one
correspondence between the true and extracted groups.

We defined a fragmented TG as one that contains the MMGs of several EGs.
Multiple situations can cause fragmentation of TGs.
In some cases, the FoF algorithm fails to recover entire TGs, selecting
instead its primary and secondary substructures (see panel
Fig.~\ref{fig:CRdef}b).
In other cases, an EG is mostly composed of galaxies from one TG, but the MMG
of another
TG is `accidentally' linked  to the first TG.
In consequence, the EG could be linked to a TG  providing only a single
member galaxy to the EG, in comparison with more members arising from
another TG\@.
When fragmentation occurred, we distinguished the \emph{primary EG}, as that
whose MMG corresponds to the MMG of the parent TG, from the other EGs, which
we called \emph{fragments}.

The dual of fragmentation is merging. In this
situation, an EG contains the MMGs of several TGs.
Proceeding similarly as for the case of fragmentation, we denoted
\emph{primary TG} of a given EG the TG whose MMG corresponds to the MMG of
that EG, denoting the other TGs as \emph{secondary}.
An example of merging is shown in Figure~\ref{fig:CRdef}c. Note that a true
group can be fragmented and its primary extracted group can be the result of
a merger of the true group with another one, as illustrated in
Figure~\ref{fig:CRdef}d.
\subsection{Local tests}
\label{sec:localtests}
Our local tests check the membership of the EGs. We defined
\emph{completeness} as the fraction of galaxies in the TG (i.e.\ within the
sphere of radius $r_{200}$) that were members of the primary EG\@. Given
this definition, it did not make sense to consider the completeness for
secondary fragments, hence we limited our tests to the primary EGs.

We defined \emph{reliability} as the fraction of galaxies in the EG that
were members of the parent TG (i.e., within the sphere of radius $r_{200}$). Here, we also
limited our tests to the primary EGs.

Mathematically speaking, these definitions of galaxy completeness, $C$, and
reliability, $R$, can respectively be written as
\begin{eqnarray}
    C=\rm \frac{TG \cap EG}{TG} \ ,\nonumber\\
    R=\rm \frac{TG \cap EG}{EG} \,. \nonumber
\end{eqnarray}

Looking at Figure~\ref{fig:CRdef}, the completeness is the fraction of
galaxies in the TG (left, green circles) recovered in the EG (right, red
circles), while the reliability is the fraction of galaxies in the EG that
belong to the TG\@.

These four quantities allow one to define the capacity of the FoF grouping
algorithm (or any other grouping algorithm) to recover groups in real space
from galaxy catalogs in redshift space.

Note that EGs that are fragments can have high reliability, while
fragmentation causes primary EGs to have reduced completeness. When EGs are
mergers of TGs, the secondary TGs lead to a decrease in the reliability, but
can have high completeness.
\subsection{Mass accuracy}
There are many properties of groups that one wishes to recover with optimal
accuracy (see Sect.~\ref{sec:intro}). We focused  here on one single
property that appeared to us as the most relevant: the group total mass. We
measured the masses of our EGs using the virial theorem formula of
\cite*{HTB85}
\begin{equation}
M_{\rm EG} = {3\pi\over G}\,\langle R \rangle_{\rm h} \,\sigma_v^2
= {3\pi\,N\over 2\,G}\,{\sum v_i^2\over \sum_{i<j} 1/R_{ij}}
\ ,
\label{MVT}
\end{equation}
where $\langle R \rangle_{\rm h} = \langle 1/R_{ij}\rangle^{-1}$ is the
harmonic mean projected separation, while $\sigma_v$ is the unbiased measure
of the standard deviation of the group velocities, given as solutions of
equation~(\ref{zfromvp}) for $v_{\rm p}^{\rm LOS}$, replacing $z_{\cos}$ by
the redshift of the MMG of the EG\@.

More precisely, we computed the accuracy of the log masses, respectively
defining the \emph{bias} and \emph{inefficiency} as the median and
equivalent standard deviation (half 16--84 interpercentile) of $\log (M_{\rm
EG}/M_{\rm TG})$, where $M_{\rm TG}$ is the mass of the TG within the sphere
of radius $r_{200}$ (see Sect.~\ref{sec:localtests}).
\subsection{Quality}
It is not simple to extract a unique pair of optimal LLs from the four tests
(fragmentation, merging, completeness, and reliability). To reduce the
number of tests, we combined fragmentation and merging into a single
\emph{global quality} and combined completeness and reliability into a
single \emph{local quality}.

We could define our qualities by multiplying $F$ (fragmentation) by $M$ (merging) and similarly, $C$
by $R$. However, one could alternatively multiply $1-F$ by $1-M$, etc. Instead, we
chose quality estimates that minimize the distance to the perfect case. The
advantage of using distance rather than multiplying probabilities is that
the former gives less weight to situations where one of the two parameters
is perfect and not the other. For example, consider the case $F=M=p$. With
the multiplication method, we would
find that $Q=p^2$ is also reached with $F=\epsilon\ll 1$, yielding $M_{\rm mult}=p^2/\epsilon$,
which can be quite large (hence plenty of merging). On the other hand, with the distance
method, we would find that $Q=p\sqrt{2}$ is also reached with $F=\epsilon\ll1$ for $M_{\rm
dist}\simeq p\sqrt{2}$, which is much more restrictive.
%
%
In a perfect algorithm, fragmentation and merging don't occur, hence $F=M=0$
they are null. We therefore chose to minimize the \emph{global quality},
defined as
\begin{equation}
Q_{\rm global} = \sqrt{F^2+M^2}
\end{equation}
%
%
Moreover, in a perfect grouping algorithm, the EGs are fully complete and
reliable, i.e. $\langle C\rangle=\langle R\rangle=1$, where the means are
over all the groups of a mass bin. We, hereafter, drop the brackets, so that
$C$ and $R$ should now be understood as means over groups within mass
bins. We then define the \emph{local quality} as
\begin{equation}
    Q_{\mathrm{local}}=\sqrt{{\left(1-
    C\right)}^2+{\left(1- R\right)}^2} \,.
\end{equation}

Both global and local qualities tend to zero for a perfect galaxy group
algorithm. So the optimal LLs will be those that minimize $Q_{\rm global}$,
$Q_{\rm local}$, mass bias and mass inefficiency.
The maximum possible value of both qualities is $\sqrt{2}$.

\subsection{Scope of the tests}
We limit our tests to TGs containing at least 3 galaxies and that are not
split by the transformations of the simulation box (see Sect.~\ref{sec:mockzspace}).
Moreover, we only consider
EGs with at least 3 galaxies and that
do not lie near the survey edges (the virial radius, 2.3 Mpc, of a
true group of log mass 15.2 in solar units, placed at $z=z_{ \rm min }=0.01$,
i.e.\ at an angle of more than $3\fdg27$) or redshift limits ($1.8\,v_{200}
\approx 2.7 \,\sigma_v$,
of the same mass group, corresponding to $3073 \, \rm km \,
s^{-1}$). Typically 60\% (sample 2) to 25\% (sample 6) of the groups are flagged (see
Appendix~\ref{sec:flag}).
Finally, the
tests of galaxy completeness and reliability, as well as mass bias and
inefficiency are restricted to primary EGs of TGs (not fragments).
\section{Results}\label{sec:results}
\begin{figure*}
    \centering
    \includegraphics[height=0.415\textheight]{%
        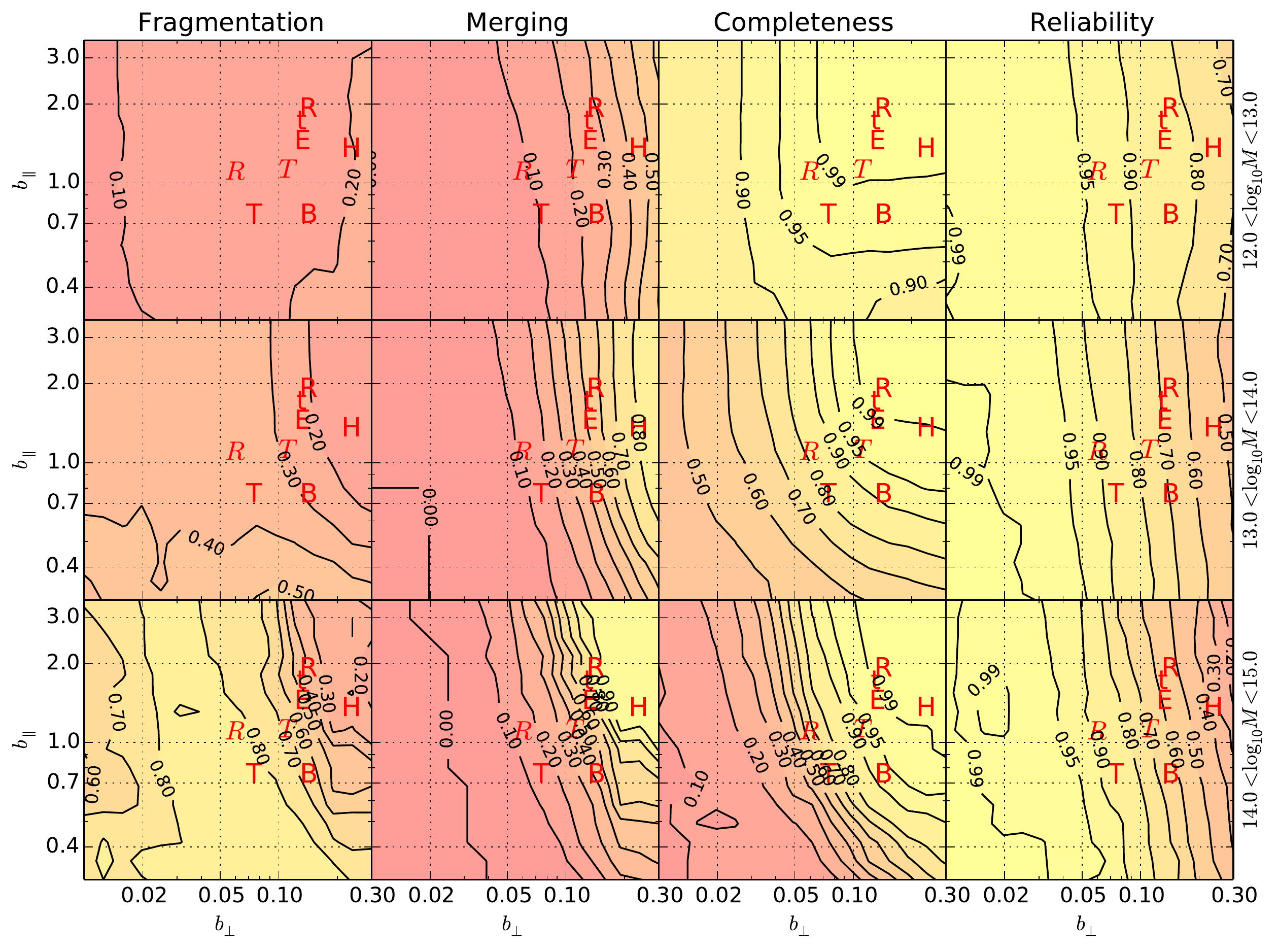%
    }
    \caption{Contours of group fragmentation (\emph{first column}) and
        merging (\emph{second column}), as well as mean galaxy completeness
        (\emph{third column}) and reliability (\emph{fourth column})
        computed for a 16$\times$16 grid of linking lengths for the nearby
        doubly complete galaxy subsample 2 in
        Table~\ref{table:samples}. Results are shown for three bins of
        true group masses,
for  unflagged groups of
        at least 3 members (for both the extracted and parent groups), and
        further restricted to primary groups in the completeness and
        reliability panels. Pairs of linking lengths corresponding
        to previous are also shown as \emph{red letters}
(H\@: Huchra \& Geller 1982;
R\@: Ramella et al. 1989;
t:        Trasarti-Battistoni 1998;
E\@: Eke et al. 2004;
B\@: Berlind et al. 2006;
T\@: Tago et al. 2010;
$R$: Robotham et al. 2011;
$T$: Tempel et al. 2014). }
\label{fig:test_true_small}
\end{figure*}
\begin{figure*}
    \centering
    \includegraphics[height=0.415\textheight]{%
        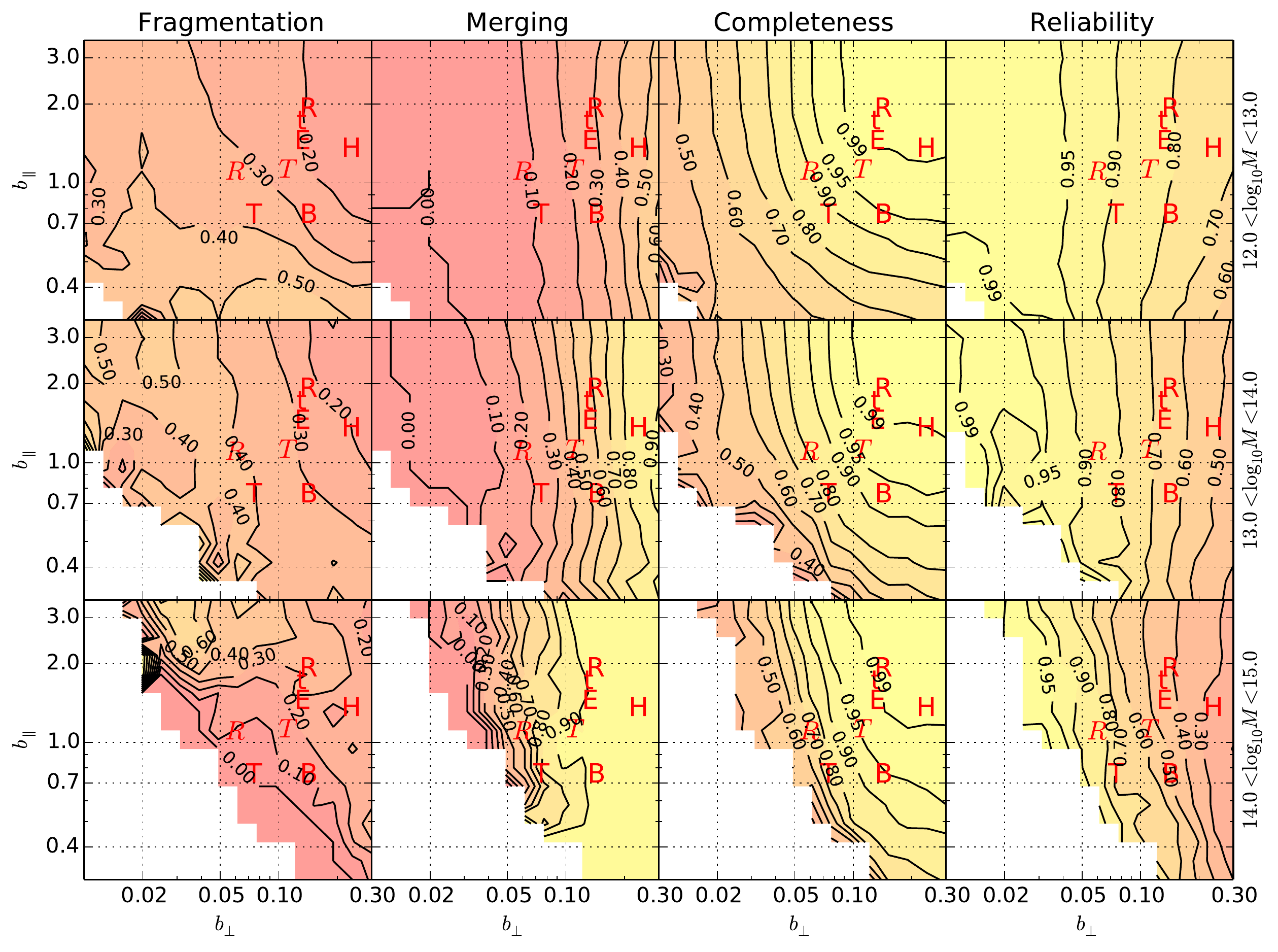%
    }
    \caption{Same as Figure~\ref{fig:test_true_small}, but
      where the
    different rows correspond to different bins of extracted group masses estimated
    from the virial theorem.
The white zones show cases where the linking lengths
        led to no unflagged groups extracted.
}
\label{fig:test_estimated_small}
\end{figure*}
\begin{figure*}
    \centering
    \includegraphics[height=0.415\textheight]{%
        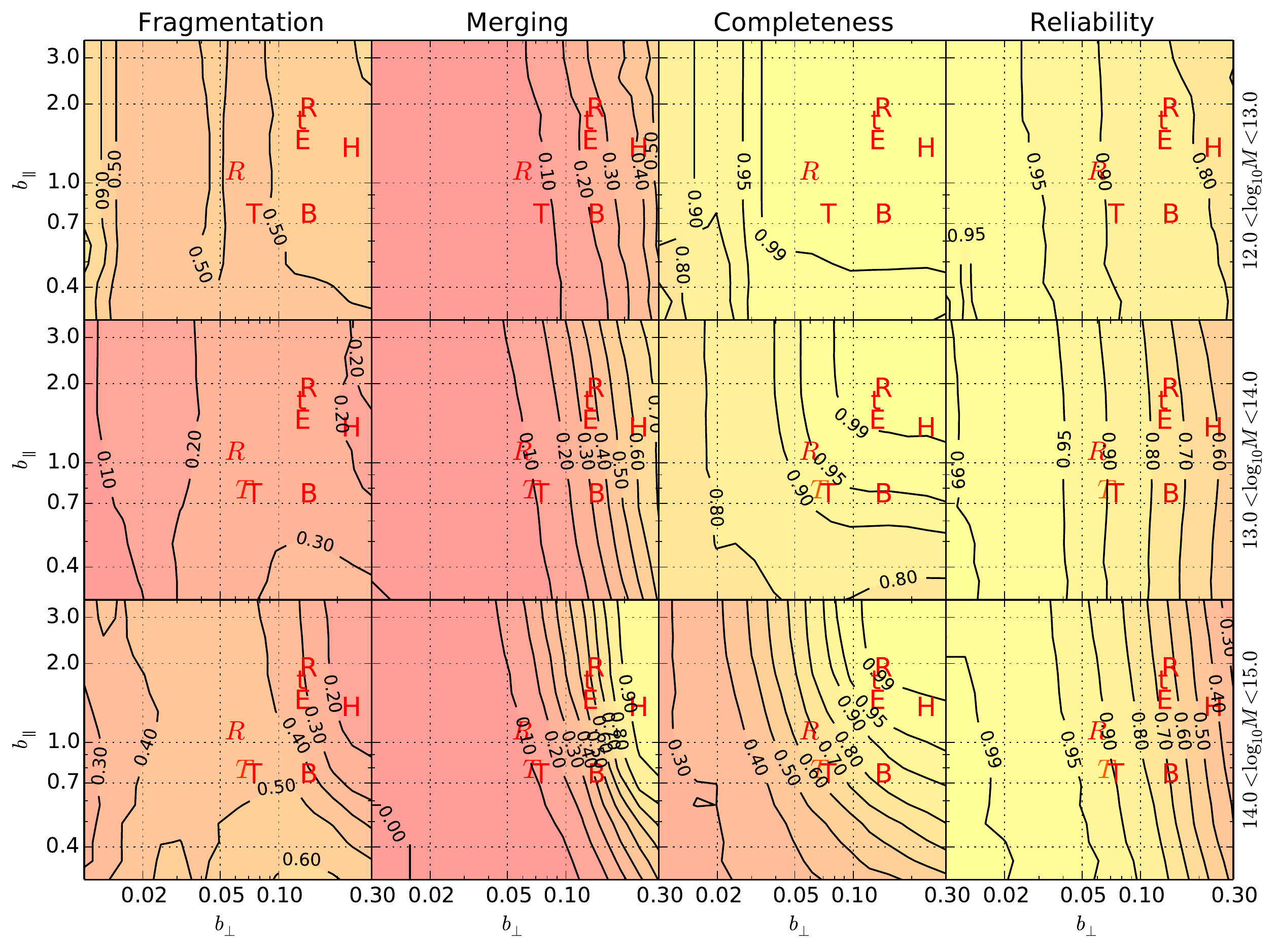%
    }
    \caption{Same as Figure~\ref{fig:test_true_small}, but for the distant
        doubly complete galaxy subsample 6 in
        Table~\ref{table:samples}.}
\label{fig:test_true_big}
\end{figure*}
\begin{figure*}
    \centering
    \includegraphics[height=0.415\textheight]{%
        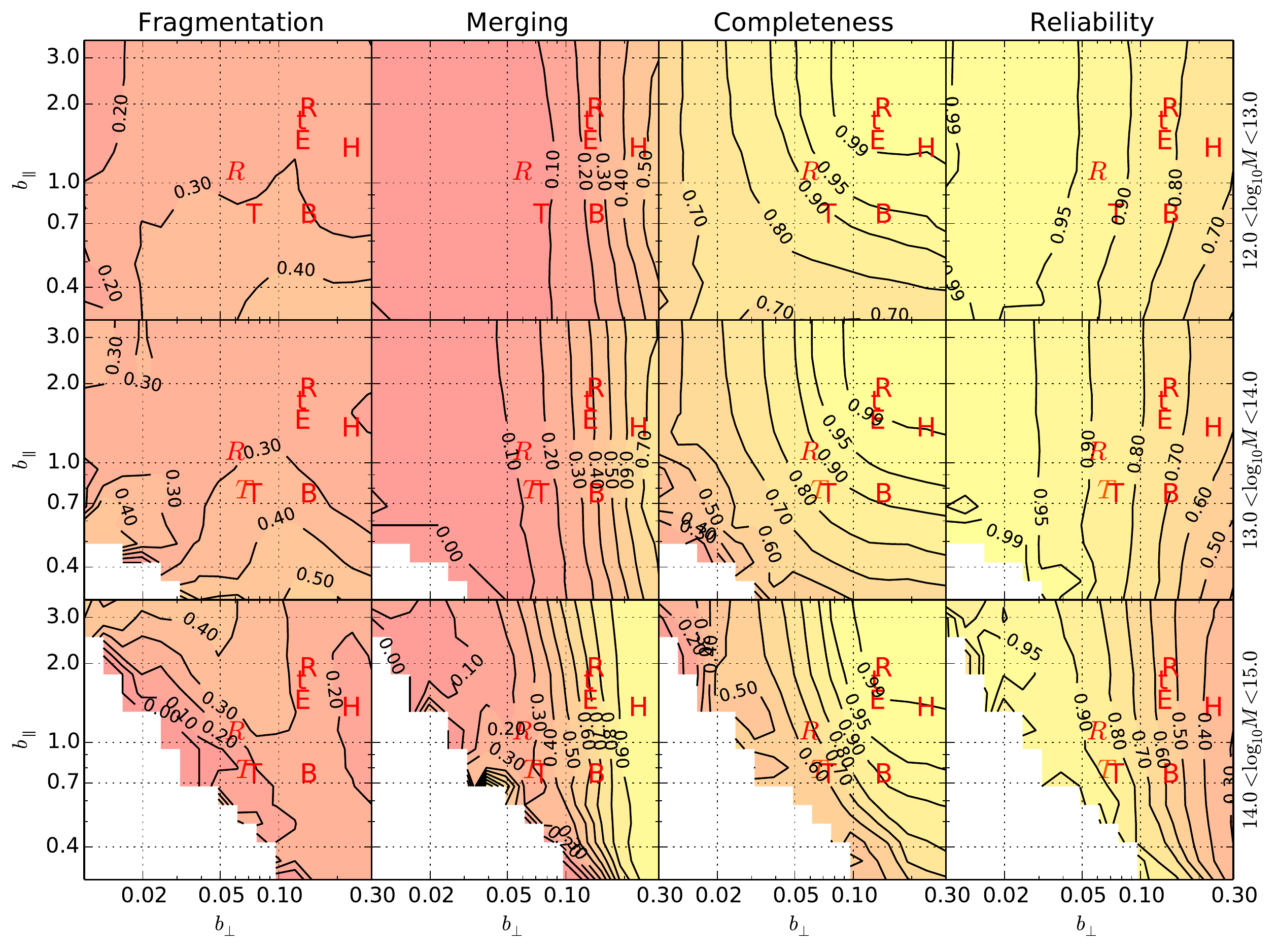%
    }
    \caption{Same as Figure~\ref{fig:test_true_big}, but where the different
    rows correspond to different bins of estimated masses.}
\label{fig:test_estimated_big}
\end{figure*}
We have applied the FoF algorithm on near and distant
doubly complete subsamples (numbers 2 and 6 in Table~\ref{table:samples}),
repeating the tests for a grid of 16$\times$16 geometrically-spaced pairs of
LLs. The results of our tests are shown in
Figs.~\ref{fig:test_true_small}--\ref{fig:masses_diff}. The LLs of the
different grouping studies listed in Table~\ref{groupalgos} are shown,
except for \cite{MZ02}, whose LLs  nearly overlap with those of \cite{Eke+04}.
\subsection{Group fragmentation and merging}
Figure~\ref{fig:test_true_small} indicates that, for the nearby doubly
complete subsample (number 2), fragmentation only affects the massive TGs (up to $\approx$80\%
of them for popular LLs), while Figure~\ref{fig:test_estimated_small} shows
that, for popular
LLs, the fragmentation is lower (10--30\%) at high EG mass, hence
fragment masses tend to be small (typically 20--40\% fragmentation at small and
intermediate estimated masses).

On the other hand, the distant doubly complete subsample behaves in almost
the opposite manner: fragmentation is most important at the lowest TG masses (roughly
50\% fragmentation, Fig.~\ref{fig:test_true_big}) and is independent of
estimated EG masses (at roughly 20--30\%, Fig.~\ref{fig:test_estimated_big}).

In any event, fragmentation tends to decrease with greater linking lengths,
as expected, although it decreases somewhat faster with increasing $b_\perp$
than with increasing $b_\parallel$.

Since merging is the dual of the fragmentation, one expects the level of
merging to vary  in the opposite way as fragmentation. Indeed,
Figures~\ref{fig:test_estimated_small} and~\ref{fig:test_estimated_big}
indicate that merging becomes more important at higher estimated
masses, respectively reaching up to 90\% and 65\% for high estimated masses
with popular choices of LLs in subsamples numbers 2 and 6. However,
Figures~\ref{fig:test_true_small} and~\ref{fig:test_true_big} shows that the
merging fraction increases only slowly with TG increasing mass, with
typically 15-40\% (increasing fast with $b_\perp$) of the TGs being merged
with other ones. Finally, merging
decreases with smaller LLs, especially with smaller $b_\perp$.
\begin{figure*}
    \centering
    \includegraphics[height=0.44\textheight]{%
        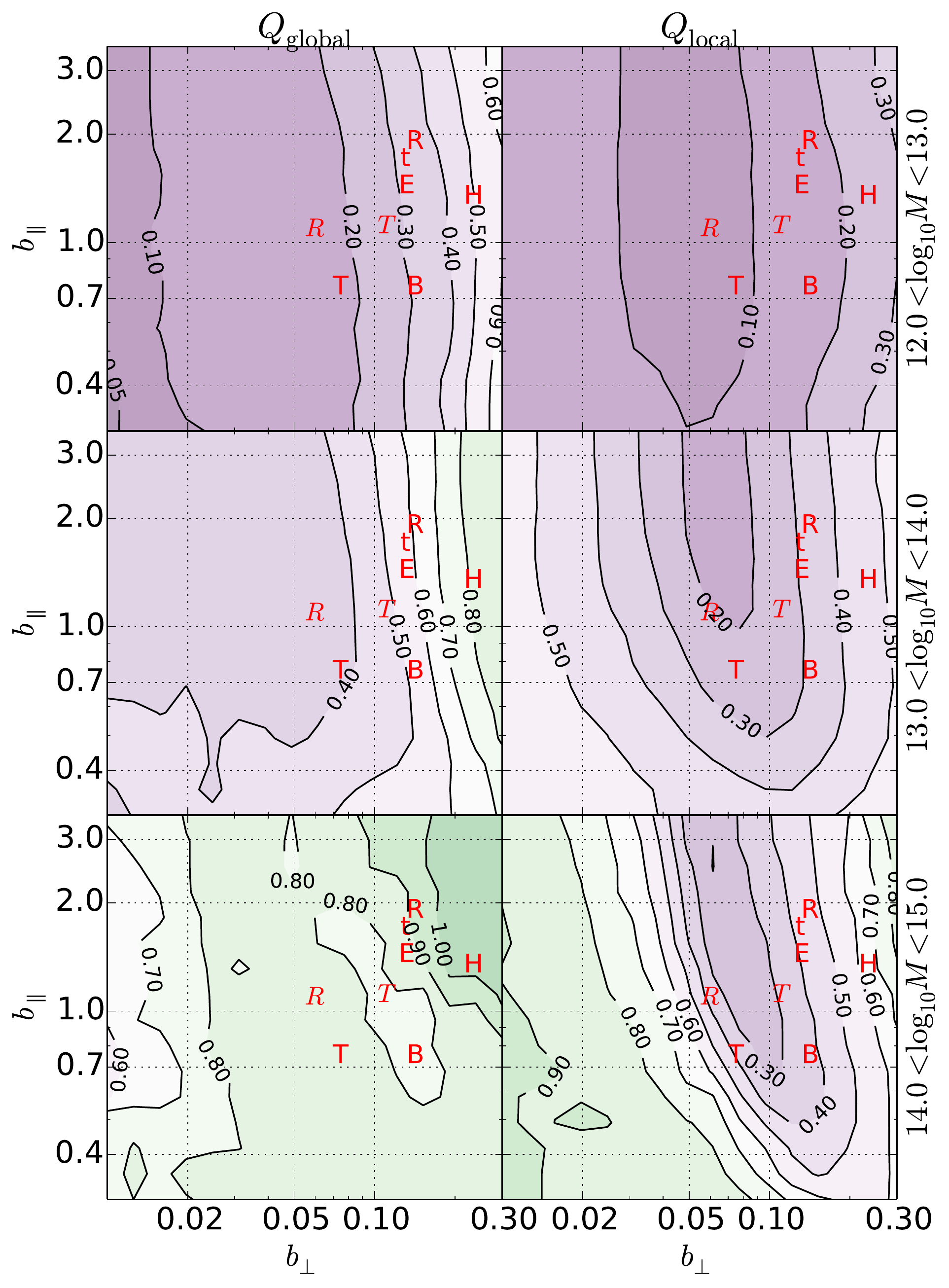%
    }
    \includegraphics[height=0.44\textheight]{%
        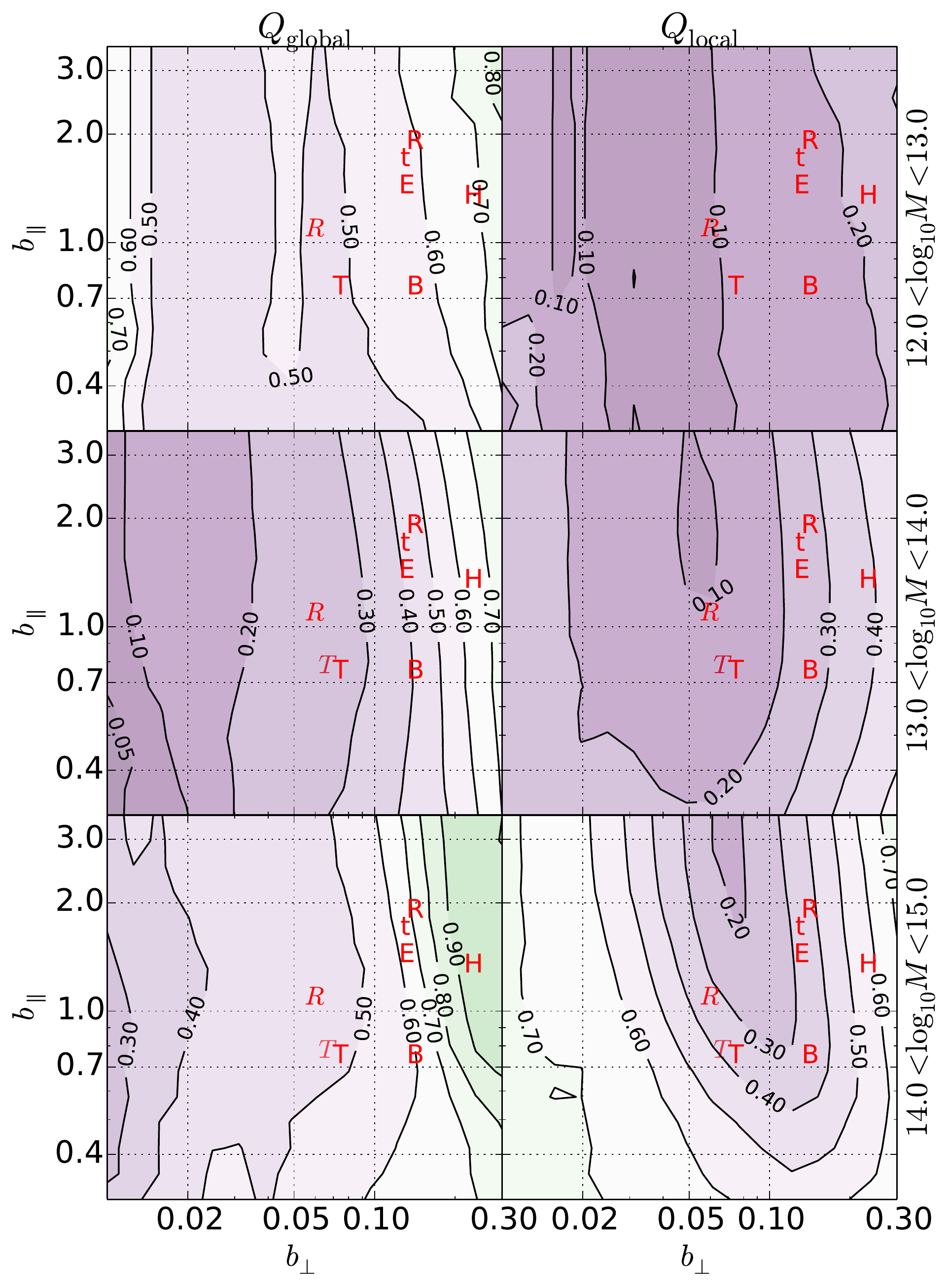%
    }
    \caption{Global and local quality factors in a 16$\times$16
        grid of linking lengths for subsamples 2 (\emph{left}) and 6
        (\emph{right}), in three bins of true masses
        Results are shown for
         unflagged groups (restricted to primary groups for $Q_{\rm
           local}$) of at least 3 members (in both the true and extracted group).
The symbols are as in   Fig.~\ref{fig:test_true_small}.
}
\label{fig:quality_true}
\end{figure*}
\begin{figure*}
    \centering
    \includegraphics[height=0.44\textheight]{%
        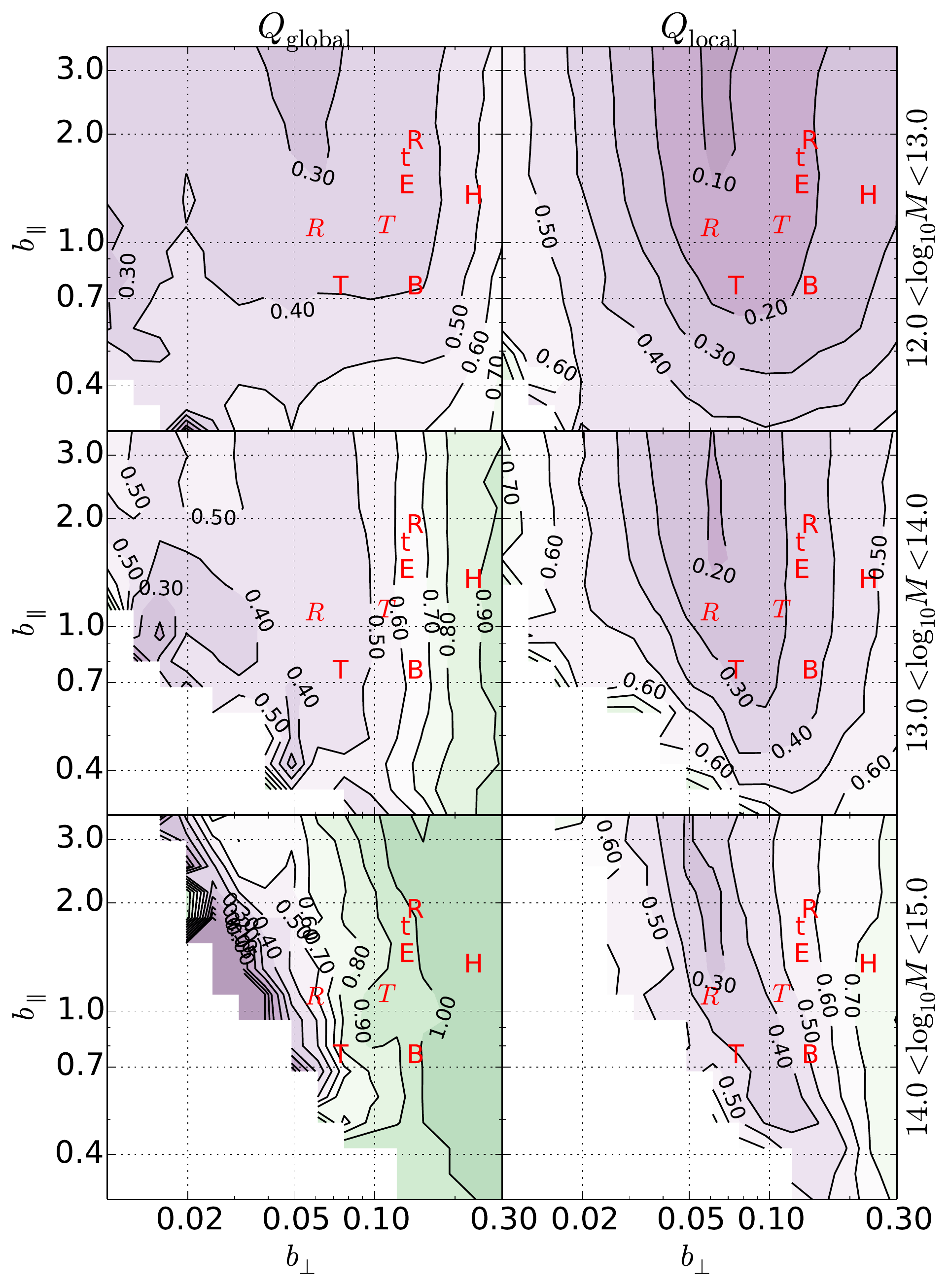%
    }
    \includegraphics[height=0.44\textheight]{%
        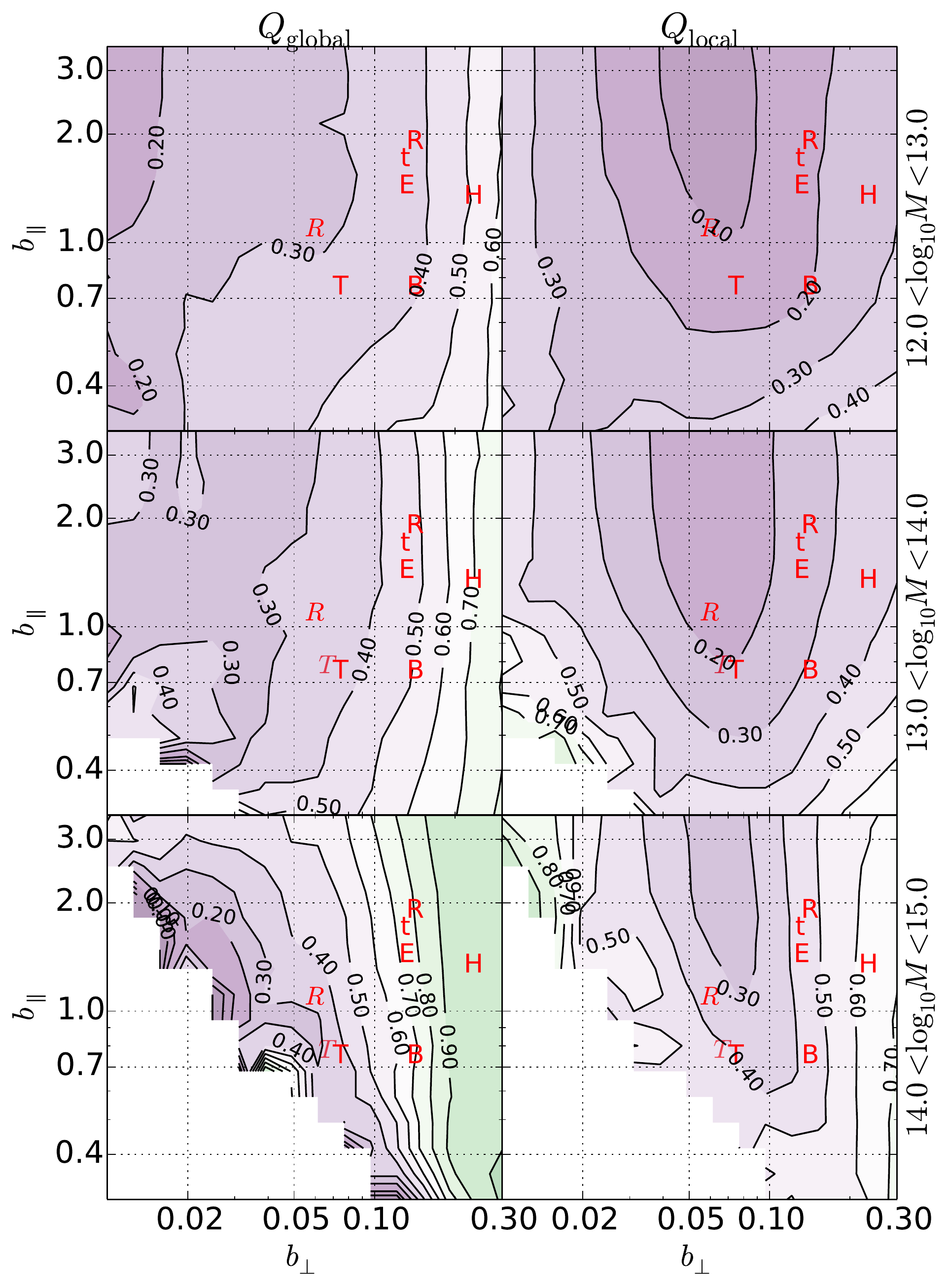%
    }
    \caption{Same as Figure~\ref{fig:quality_true} but in bins of
    estimated masses.
The white zones show cases where the linking lengths
        led to no unflagged groups extracted.
    }
\label{fig:quality_estimated}
\end{figure*}

Figures~\ref{fig:quality_true} and~\ref{fig:quality_estimated} show the
$Q_{\rm global}$ quality indicator that combines fragmentation and merging
into a single parameter. These figures show that decreasing $b_\perp$ leads
to a better tradeoff between fragmentation and merging, i.e.\ that the
decrease of merging with decreasing $b_\perp$ has a stronger effect than the
increase of fragmentation with decreasing $b_\perp$: the optimal $Q_{\rm
  global}$ is often reached for $b_perp < 0.02$.
\subsection{Galaxy completeness and reliability}
Figures~\ref{fig:test_true_small} and~\ref{fig:test_true_big} indicate that
completeness is very high ($>99\%$) at low TG masses, and decreases to lower
values ($60-99\%$) at high TG mass. A weaker trend occurs when EG mass
is substituted for TG mass (see Figs.~\ref{fig:test_estimated_small}
and~\ref{fig:test_estimated_big}).
Since high mass TGs are less complete, their estimated masses should be
smaller, and the EGs with high masses will be the lucky complete ones, which
explains the weaker trend of completeness with EG mass.
Note that we are only considering primary groups of at least 3 members.
The transverse and LOS linking lengths have roughly the
same impact on galaxy completeness.

The reliability of the group membership decreases with increasing EG mass
(Figs.~\ref{fig:test_estimated_small} and~\ref{fig:test_estimated_big}):
regardless of the subsample, the reliability is 80--90\% for low mass EGs,
but only 50--85\% for high mass EGs.
The value of $b_\parallel$ has virtually no effect on galaxy
reliability. 
We will discuss this lack of convergence of the reliability with
$b_\parallel$  in Sect.~\ref{sec:discussion}.

Galaxy reliability also decreases with the masses of the TGs, but the trend
is weaker (Figs.~\ref{fig:test_true_small} and~\ref{fig:test_true_big}):
as the reliability decreases from 85--95\% to 60--90\%, roughly independent
of the subsample.

The right panels of Figures~\ref{fig:quality_true}
and~\ref{fig:quality_estimated} show that, again, the transverse LL appears
to be more decisive than the LOS one when combining galaxy completeness and
reliability into a single local quality factor.
%

%
%
\subsection{Mass accuracy}
\begin{figure*}
    \centering
    \includegraphics[height=0.44\textheight]{%
        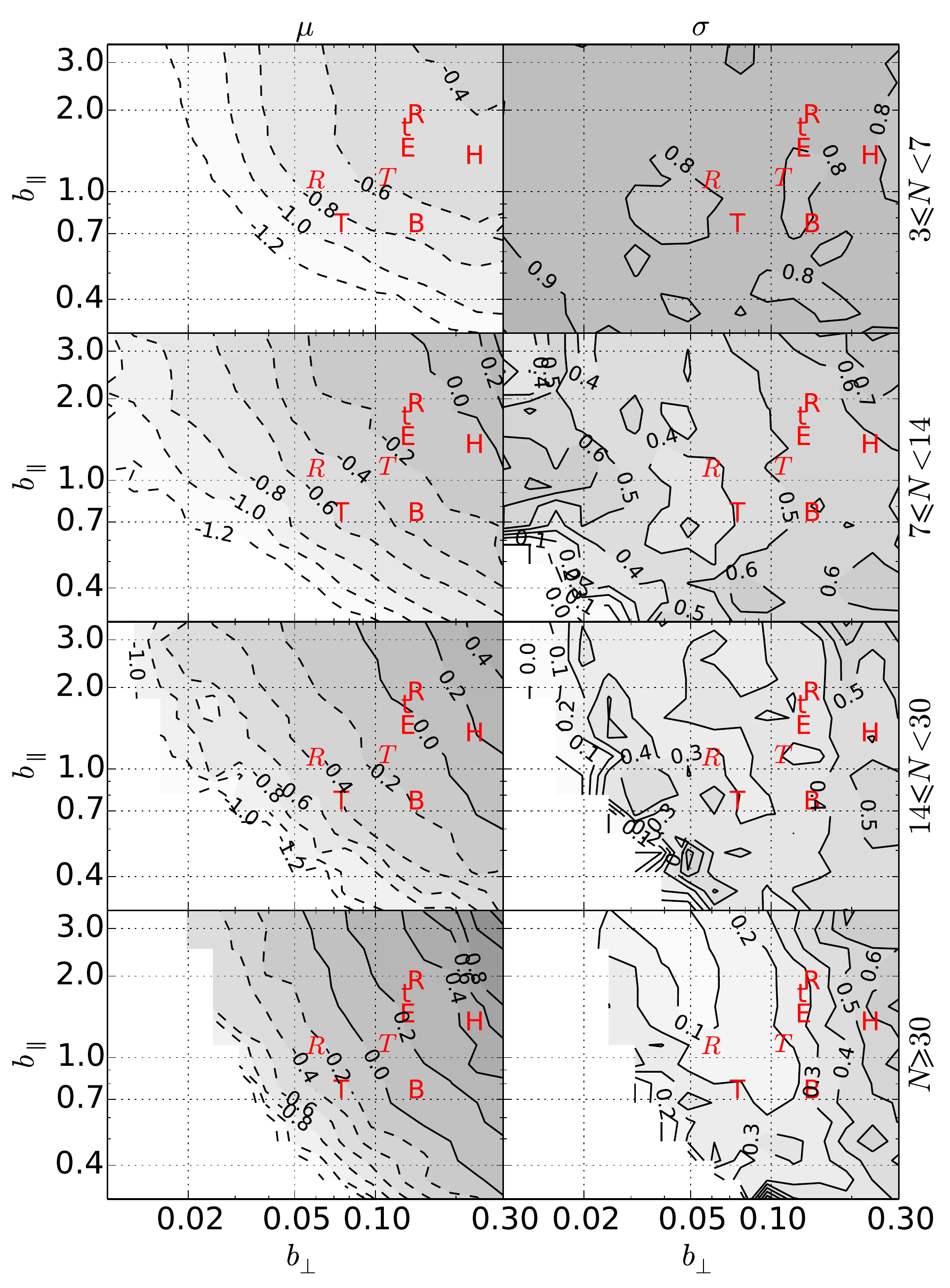%
    }
    \includegraphics[height=0.44\textheight]{%
        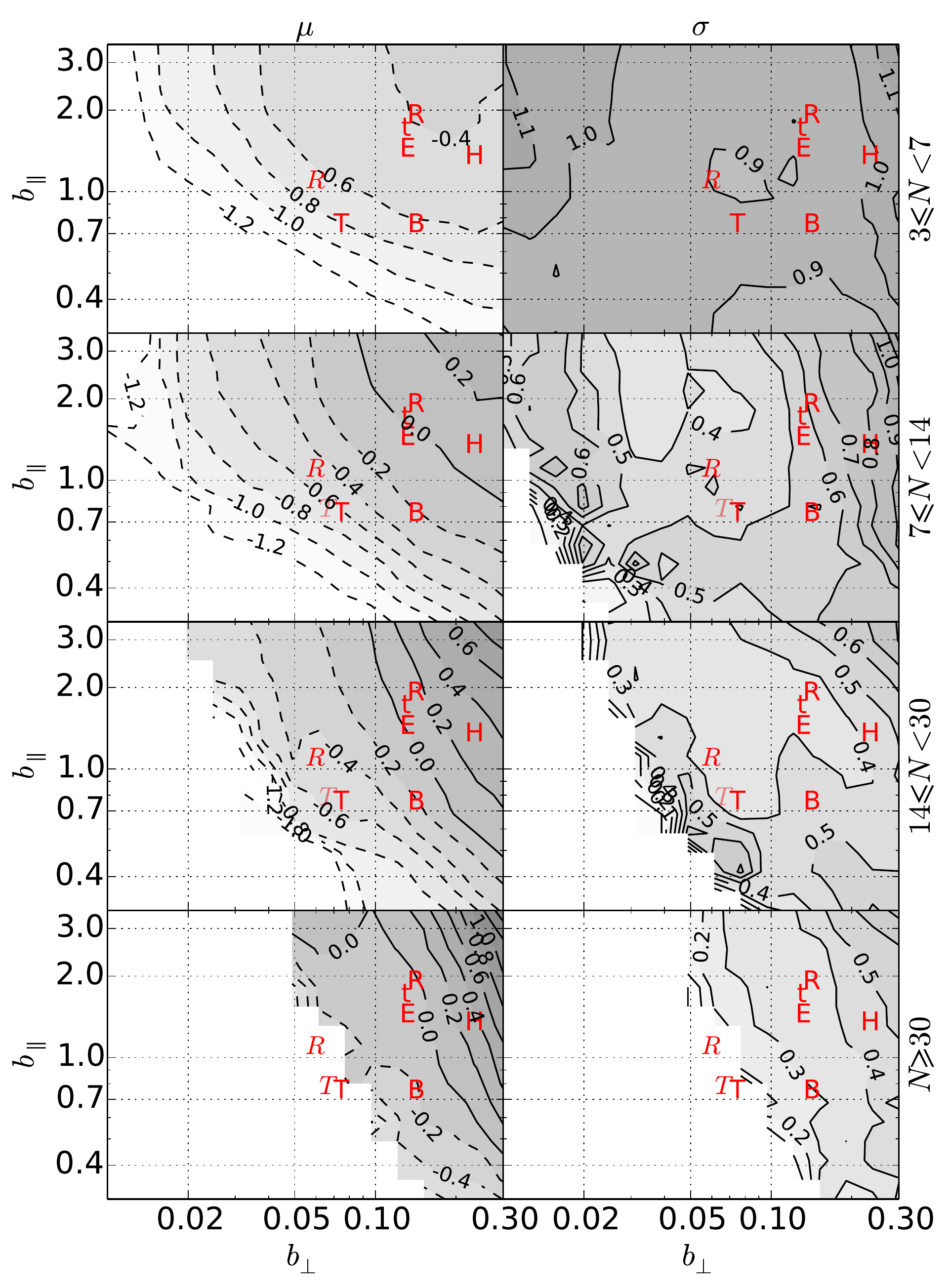%
    }
    \caption{Bias ($\mu$) and inefficiency ($\sigma$) of the group masses
      estimated  by the virial theorem (eq.~[\ref{MVT}]) on our
        16$\times$16  grid of linking lengths, in four bins of extracted
        group richness (we do not consider extracted groups for which the
        parent true group has $\leq3$ members). The bias and inefficiency are respectively computed
        as the median and half 16--84 interpercentile of
        $\log_{10}\left(M_{\rm EG}/M_{\rm TG}\right)$. Results are shown for
        primary, unflagged groups. The left and right panels are
        respectively  for galaxy  subsamples 2 and 6. 
The symbols are as in   Fig.~\ref{fig:test_true_small}.
The white zones indicate linking lengths with no unflagged groups extracted.
    }
\label{fig:masses_diff}
\end{figure*}

The left columns  of the two panels of Figure~\ref{fig:masses_diff} show
that the primary EG masses recovered by the FoF algorithm are systematically biased
low: for the popular choices of LLs, the bias ($\mu$) is as strong as $-0.6\pm0.2$
dex at low multiplicity ($N_{\rm EG}\leq 6$), decreasing to $0.0\pm0.3$ dex at
high multiplicity ($N_{\rm EG}\geq 30$).

The right columns of the two panels of Figure~\ref{fig:masses_diff} indicate
that, even if the biases could be corrected for, the masses cannot be
recovered to better than 0.8--0.9 dex at low multiplicity, improving to 0.2
dex at high multiplicity. 
The  inefficiency
($\sigma$) is minimal for $b_\perp \approx 0.05$ (within a factor 2) and
$b_\parallel \approx 1.0$ (low richness) or $b_\parallel \ga 1.0$
(intermediate and high richness).
For transverse LLs within 40\% of $b_\perp=0.1$, the
inefficiency is not very insensitive to $b_\parallel$.

The situation becomes even worse when fragments are included in the
statistics. In this work, we have separated the accuracy of the group masses
with the occurrence of group fragmentation. But observers cannot tell if a
group is a fragment or a primary EG.

%
%
\section{Conclusions and Discussion}
\label{sec:discussion}

Before testing the FoF algorithm using a mock galaxy catalog in redshift
space, we first argued on physical grounds (Sect.~\ref{sec:fofpred}) that the
normalized transverse linking length, ought to be $b_\perp \approx 0.10$
(slightly increasing with richness) to extract 95\% of the galaxies within
the virial radius of NFW true groups.  We also argued that, restricting the
galaxies along the line-of-sight to $\pm1.65\,\sigma_v$ (95\% of the
galaxies) for groups defined to be 200 times denser than the critical density
of the Universe, requires $b_\parallel/b_\perp \approx 11$, hence
$b_\parallel \simeq 1.1$.  These LLs are estimated from our mocks that are
based upon the Millennium-II simulation that had adopted $\Omega_{\rm
  m}=0.25$. Converting to $\Omega_{\rm m}=0.3$ yields $b_\perp=0.11$ and
$b_\parallel=1.3$.  Finally, estimating the contamination by interlopers, we
predict between 80\% (NFW model extended outwards) to 90\% (NFW model
truncated to sphere plus random interlopers) galaxy reliability.

We then built a mock redshift space galaxy catalog with the properties of the
flux-limited SDSS primary spectroscopic sample, from which we extracted 2
subsamples that are doubly complete in distance and luminosity
(Sect.~\ref{sec:mock}).  We then extracted groups from both of these
subsamples, running the standard FoF algorithm for $16\times16$ pairs of
linking lengths.  In each case, we measured the fraction of true groups that
were fragmented in the FoF extraction process, the fraction of extracted
groups that were built by the merging of several true groups, as well as the
bias and inefficiency with which the group masses were extracted. Moreover,
we computed the completeness and reliability of the galaxy membership
relative to the spheres of radius $r_{200}$ in which the true groups are
defined.

We analyzed group fragmentation, merging, galaxy completeness and
reliability, mass bias and inefficiency for two doubly complete subsamples
and in bins of true and estimated mass or estimated richness (for the mass
accuracy).

We found that massive true groups are more prone to fragmentation, as
expected, but that, for popular choices of linking lengths, the probability
of fragmentation is greatest (30\%) at low estimated mass, i.e.\ the
fragments are of low mass.
The process of fragmentation of rich (massive) groups  is similar to images
of large galaxies being preferentially fragmented by automatic image
extraction pipelines (e.g., \citealp{DePropris+07}).

Group merging is low at low estimated mass, but increases drastically
to reach 40--90\% (for popular linking lengths) at high estimated
mass. Galaxy completeness is high, typically $>80\%$. Galaxy reliability is
typically 75 to 90\% depending on group mass..

Our analytical prediction of 95\% completeness for $b_\perp\simeq 0.10$ is
only met for groups of high true masses (Figs.~\ref{fig:test_true_small} and
\ref{fig:test_true_big}). Groups of low mass will have more concentrated
galaxy populations, which will lead to smaller values of ${\rm
  Max}(S_\perp)/r_{200}$, hence smaller values of $b_\perp$.
Also, our analytical prediction of 80--90\% reliability for groups with
$b_\perp=0.10, b_\parallel=1.1$ is accurate for groups of all masses of the
distant subsample (Fig.~\ref{fig:test_true_big}). However, for the nearby
subsample (2), our predicted reliabilities are only \nobreak{accurate} for groups of
low true masses, 
but optimistic for higher mass groups, for which $R \simeq 70-75\%$.

Group merging and galaxy reliability depend little on $b_\parallel$,
especially at high transverse linking length, $b_\perp > 0.1$, where the
galaxies are extracted to projected radii beyond $r_{200}$, hence the
contamination by interlopers is mainly in the transverse direction.
The lack of optimal $b_\parallel$ for galaxy reliability may seem surprising
at first. We checked our analysis by measuring the reliability for
$b_\perp=0.1$, for a very wide range of $b_\parallel$ extending from 0.3 to 40.
%
\begin{figure}
    \includegraphics[width=\linewidth]{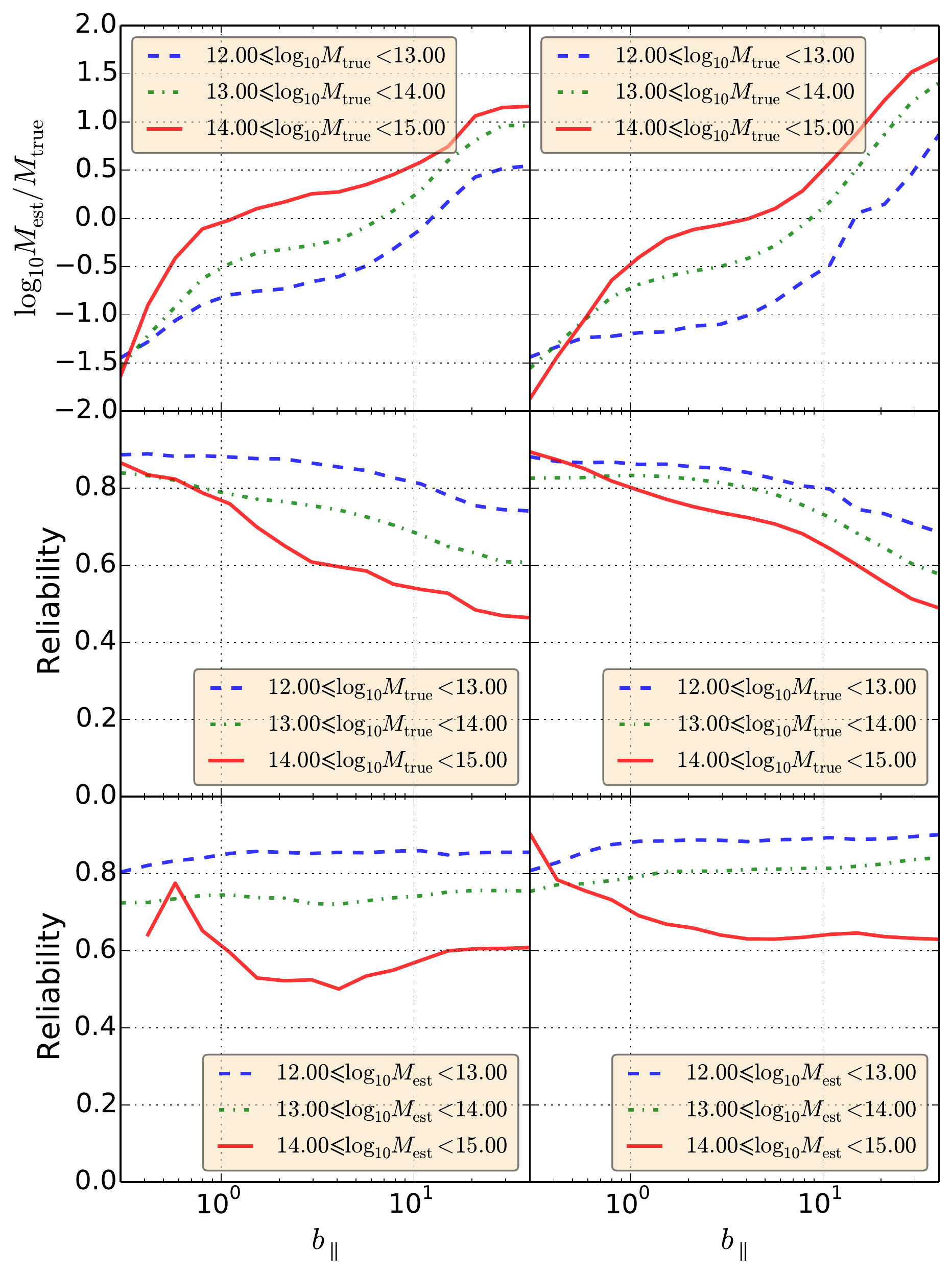}
    \caption{Variation of the mass bias and reliability as a function of
     $b_\parallel$ for $b_\bot =0.1$, for subsamples 2 (\emph{left}) and 6
      (\emph{right}).} 
\label{fig:parallel_behaviour}
\end{figure}
The  top panels of Figure~\ref{fig:parallel_behaviour} indicate that the
reliability does end up decreasing fairly fast beyond some large value of
$b_\parallel \simeq 6$, i.e.\ beyond the limits of
Figures~\ref{fig:test_true_small} and~\ref{fig:test_true_big}. The second
row of panels of  Figure~\ref{fig:parallel_behaviour} show a different
behavior in bins of estimated mass. This is the consequence of the estimated
mass increasing very fast with $b_\parallel$, as shown in the bottom panels
of  Figure~\ref{fig:parallel_behaviour}. The increase, with increasing
$b_\parallel$, of the mass bias is roughly parallel to the corresponding
decrease of the reliability (in bins of TG mass). At low $b_\parallel$, the
reliability decreases fairly rapidly and the mass bias increases rapidly
(towards zero), then both settle into an almost constant plateau in the
range $1.4 \la b_\parallel \la 8$, then both worsen rapidly up to
$b_\parallel\simeq 25$, beyond which both saturate, because the longitudinal
link is so large that one reaches the minimum and maximum redshifts of the
subsample, where most groups are flagged. Massive groups that are built from
TG merging can be fairly reliable if the secondary TGs have negligible mass
relative to the primary one. This explains why $R$ remains fairly high when
$M$ is high. The plateau around \bpar$\approx 3$ appears to represent the
range of optimal longitudinal LLs.

\begin{figure}
    \centering
    \includegraphics[width=0.7\linewidth]{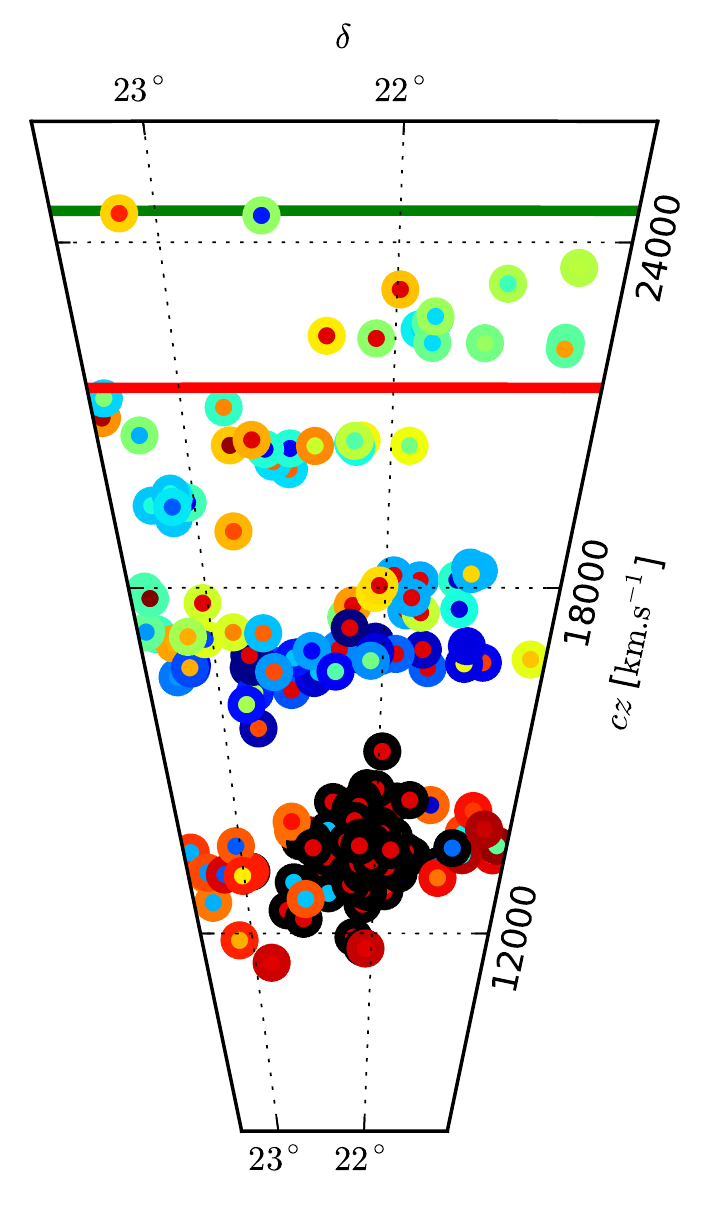}
    \caption{An example of group and halo for $b_\parallel=20.8$ and
        $b_\bot=0.1$ for subsample 4. The width of the cone is  exaggerated
        by a factor of roughly 5 for illustrative purposes. \emph{Outer} and
        \emph{inner circle colors} respectively refer to the TGs and EGs.
        The \emph{horizontal green} and \emph{red lines} respectively
        indicate the maximum redshift, $z_{ \rm max }$ and the redshift
        where galaxies are flagged for being close to $z_{ \rm max }$. Some
    galaxies of the red EG, whose TG is the black one, are flagged for being
close to $z_{ \rm max }$, hence the group would not be considered in our
tests.}
\label{fig:group}
\end{figure}

An illustration is given in Figure~\ref{fig:group}, where a given EG has
reached the limits of the catalog with a very large value of \bpar{}.
Fig.~\ref{fig:group} also shows that interloping TGs are highly clustered.
This may explain why increasing $b_\parallel$ has only a small effect on
galaxy reliability: there is a void behind the main TG (black outer
circles).

While fragmentation, measured in bins of true group mass, decreases with
increasing $b_\parallel$, as expected (Figs.~\ref{fig:test_true_small} and
\ref{fig:test_true_big}), we find that in bins of estimated
mass, the fraction of groups that are (secondary) fragments increases with
$b_\parallel$ (Figs.~\ref{fig:test_estimated_small} and
\ref{fig:test_estimated_big}). We believe that this is caused by interlopers
increasing the 
group estimated mass (Fig.~\ref{fig:parallel_behaviour}).

The masses, estimated with the virial theorem (eq.~[\ref{MVT}]) are a strong
function of the multiplicity of the extracted group. The estimated masses
are systematically biased low, especially for low extracted group
multiplicities (typically by a factor 4!). Similar trends have been found for
FoF groups \citep{Robotham+11} and for other, mostly dynamical,
group mass estimators \citep{Old+14}.
The estimated group masses are inaccurate, even after correcting for the
biases: the typically errors are 0.8--0.9 dex at low multiplicity,
decreasing to 0.3 dex at high multiplicity.

The optimal completeness and reliability of the galaxy membership lead to
fairly extreme linking lengths, i.e. $b_\perp < 0.1$ and $b_\parallel > 2$.
However, the use of such a small transverse linking length amounts to
extracting the inner regions of groups, thus missing their outer envelopes.
Indeed, one notices that fragmentation worsens at increasingly lower values
of $b_\perp$. Therefore, our attempt to define a local quality by combining
galaxy completeness and reliability is of little use if one wishes to
recover galaxies out to close to the virial radii of groups.

In fact, the optimal linking lengths depend on the scientific goal:
\begin{itemize}
\item statistical studies of environmental effects require high reliability
  (say $R>  0.9$), accurate masses and, to a lesser extent, minimal fragmentation.
\item cosmographical studies of group mass functions require accurate masses, minimal group
  merging and fragmentation.
\item studies for followups at non-optical wavelengths (e.g. X-rays), benefit
  from high completeness.
\end{itemize}

For statistical studies of environmental effects, it seems best to adopt
$b_\perp \simeq 0.06$, $b_\parallel \approx 1.0$, for which the reliability is
roughly as high as it gets for the choice of $b_\perp$: over 90\% at low
$M_{\rm EG}$ and over 80\% at intermediate and high $M_{\rm EG}$. Then,  the
completeness is higher than 70\% at high estimated mass and much higher at
low $M_{\rm EG}$.
The mass inefficiency is minimal, but with this choice of LLs, there will be
virtually no EGs with more than 30 galaxies in the distant more luminous
subsample
(Fig.~\ref{fig:masses_diff}).

This choice of LLs is close to that of \cite{Robotham+11}, which may seem
obvious since both studies used some form of optimization of the
LLs. However, the details of the optimization criteria are somewhat
different: \citeauthor{Robotham+11} multiplied four criteria: basically the
group completeness and reliability, which bears some resemblance to our group
fragmentation and merging, but theirs is based on TG-EG pairs that have more
than half their galaxies in common, as well as two measures of a combination
of galaxy completeness and reliability, averaged over TGs and EGs
respectively.
Our analysis differs in that we directly constrained group fragmentation and
merging, as well as galaxy completeness and reliability for primary
fragments, and finally mass accuracy.

For cosmographical and other studies involving accurate group mass
functions, it appears best to adopt $b_\perp \simeq 0.05$, $b_\parallel
\simeq 2$, as lower $b_\parallel$ increases fragmentation
(Figs.~\ref{fig:test_estimated_small}
and~\ref{fig:test_estimated_big}), while higher $b_\parallel$ causes too
high group fragmentation at high EG masses. This value of $b_\parallel\simeq 2$ is in agreement
with the intersection of the regions of $(b_\perp,b_\parallel)$ space  that
optimize both the multiplicity function  and velocity dispersions obtained by
\cite{Berlind+06}.

Finally, for non-optical followups, for which galaxy completeness is perhaps
the sole important parameter, one should privilege large linking lengths,
e.g. $b_\perp \simeq 0.2$, $b_\parallel \simeq 2-4$. However, one can also
adopt $b_\perp = 0.1$, $b_\parallel \simeq 2-4$, for which the completeness
is
greater than 95\% at all masses and for both subsamples.

Converting from
 $\Omega_{\rm m} = 0.25$ (Millennium-II Simulation) to
 $\Omega_{\rm m} = 0.3$ (WMAP-Planck compromise),
$b_\perp$ must be increased by 6\% (eq.~[\ref{bperpfromDelta}]) to
$b_\perp \simeq 0.07$ for the choices optimizing environmental or
cosmographical studies. Since $b_\parallel/b_\perp$ is independent of
$\Omega_{\rm m}$ at given $\Delta$, \bpar{} must also be increased by 6\%,
i.e.\ to \bpar$\approx1.1$ for environmental studies.

We finally note that while high estimated mass group fragmentation  and
merging depends on the particular doubly complete subsample, galaxy
completeness and reliability as well as mass accuracy depend little on the
subsample.~\cite{Berlind+06} had similarly concluded that the doubly complete
subsample influenced little their tests of the group
multiplicity function and the accuracy of projected radii and velocity dispersions.

FoF grouping techniques can be used as a first guess for other more refined
grouping methods
\citep{YMvdBJ05,Yang+07}.
In a future paper \citep{DM14b}, we will present another grouping algorithm,
which is not an FoF, but is instead a probabilistic grouping algorithm that
is built upon our current knowledge of groups and clusters (partly from
X-rays and independent of FoF analyses of optical galaxy samples) and
from cosmological $N$ body simulations.

\section*{Acknowledgments}
We thank the referee, Jon Loveday, for his careful reading of the manuscript
that improved this article. We also acknowledge Darren Croton for a useful
discussion. 
The Millennium-II Simulation database used in this paper and the web
application providing online access to them were constructed as part of the
activities of the German Astrophysical Virtual Observatory (GAVO).
We are grateful to Michael Boylan-Kolchin and Qi Guo for respectively allowing
the outputs of the  Millennium-II simulation and the Guo semi-analytical
model to be available to the public, and Gerard Lemson for maintaining the
GAVO database.

%
%
\bibliography{references}
%


\appendix
\section{Galaxy search}\label{app:grid}
\begin{figure}
 \centering
 \includegraphics[width=\linewidth]{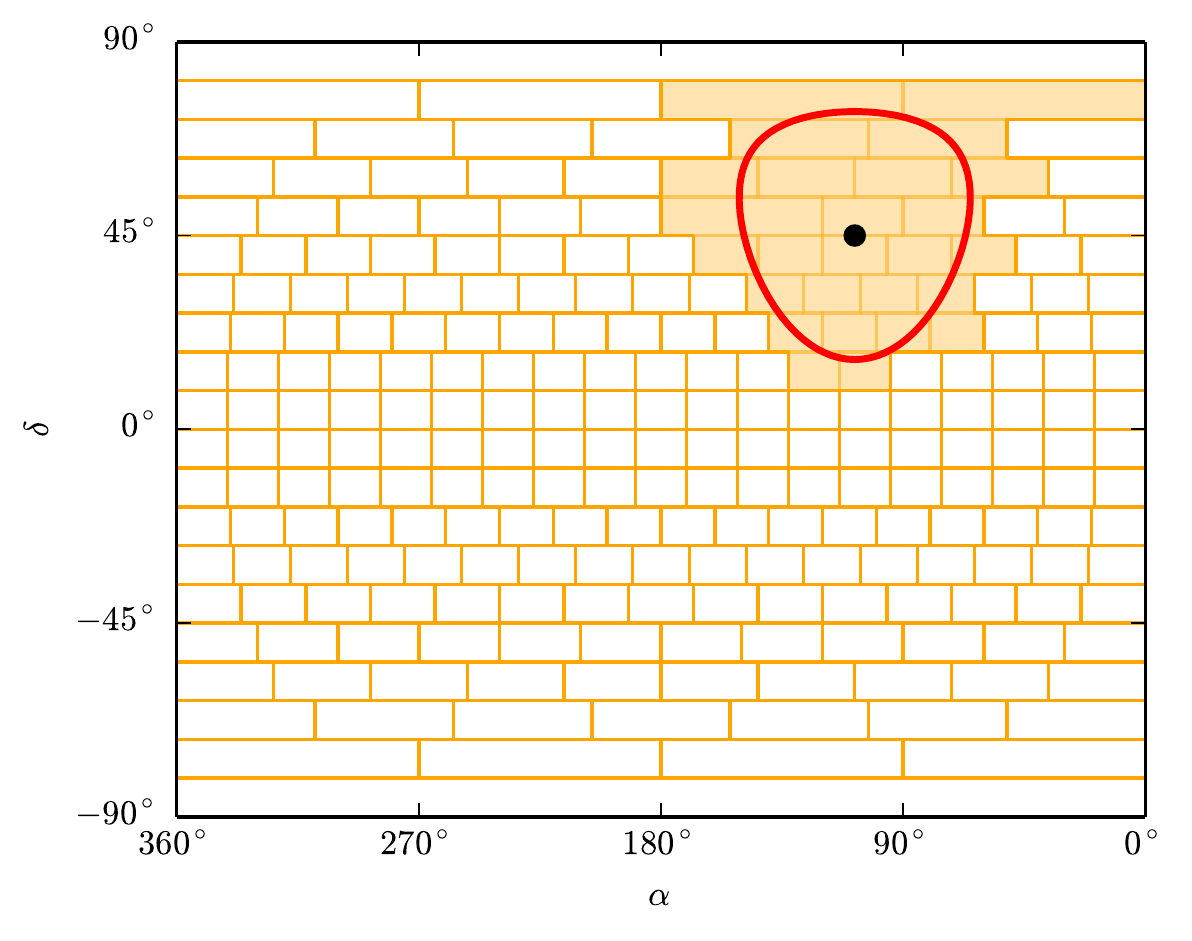}
 \caption{An illustration of the grid on the celestial sphere for a fast
   search of neighbors.
 Selected boxes to search are highlighted, for the given angular
distance from the central point (\emph{red line}). Note that both the search
angle and the cell size
are greatly exaggerated for illustrative purposes.
}
\label{fig:mesh}
\end{figure}
Implementing galaxy grouping algorithms, such as  FoF, requires the search
for galaxy neighbors, which can  be very time consuming if one computes all
$N(N-1)/2$ separations between the $N$ galaxies.
We proceed in two steps, first selecting galaxies meeting the transverse
link, then restricting these galaxies to those that also meet the LOS link.
We built a two-dimensional grid on the sky coordinates with constant steps
in declination and steps proportional to $1/\cos \delta$ in right ascension
so that the length in right ascension (at the mean declination of the band
of cells) is roughly equal to the step in declination.
For each galaxy, we determine the cells that require searching for
neighbors, and then we search using spherical trigonometry relations (see an
illustration of this method in
Fig.~\ref{fig:mesh}).
%
The LOS
link is then checked (without subdividing into LOS cells).

The computer time required to build the FoF groups is substantially reduced
compared to the brute-force computation between pairs.
The bottleneck of our tests involves the computation of the harmonic mean
radius when measuring the EG mass by the virial theorem (eq.~[\ref{MVT}]).

\section{Fraction of flagged  groups}
\label{sec:flag}
\begin{figure}
    \includegraphics[width=0.95\linewidth]{%
        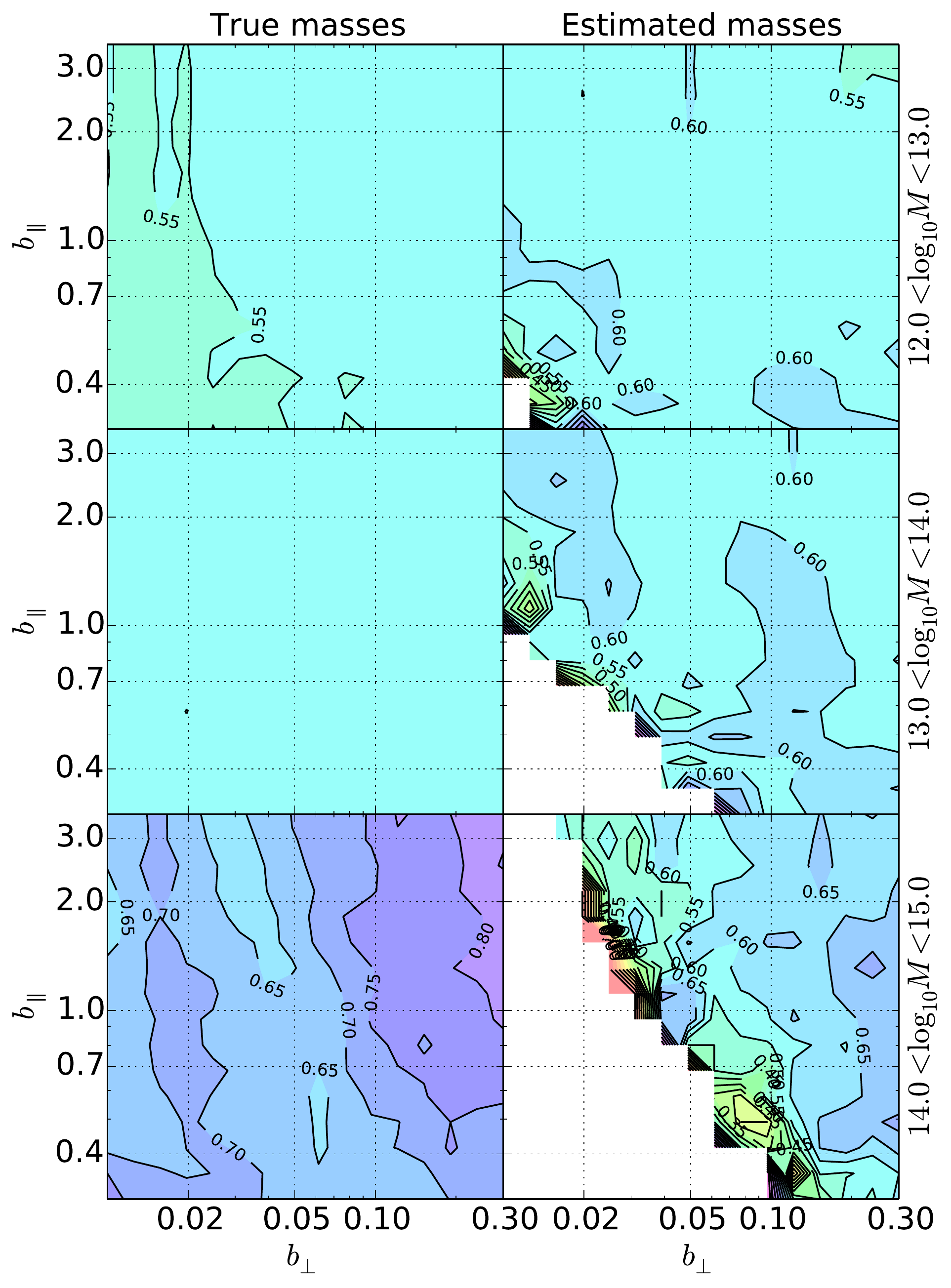%
    }
    \includegraphics[width=0.95\linewidth]{%
        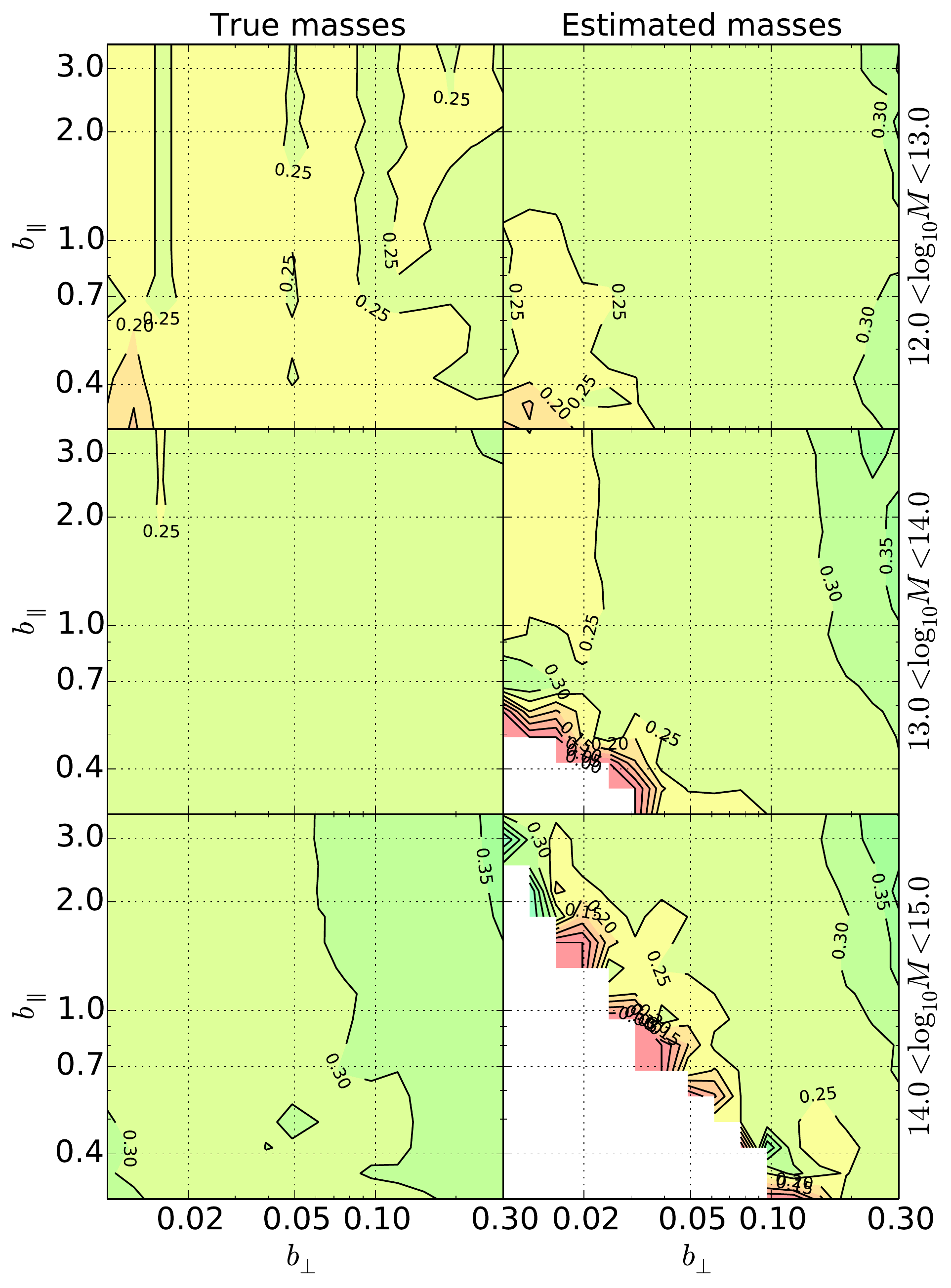%
    }
    \caption{Fraction of selected groups flagged as either split by the transformations
      of the simulation box or lying close to the edges of the
        mock galaxy survey for catalogs 2 (\emph{top}) and 6
        (\emph{bottom}), in bins of true and estimated masses.
}
\label{fig:fractions}
\end{figure}

Figure~\ref{fig:fractions} displays the fraction of flagged groups, either
because their parent groups were split in the simulation box transformations
(Sect.~\ref{sec:mock}) or because they are close to the survey edges and redshift limits.
The fractions of flagged galaxies are greater in the nearby samples, because
the survey edges and redshift limits are more important in this smaller
volume sample.

\end{document}